\documentclass[twocolumn]{aastex61}

\usepackage{natbib}
\usepackage{comment}
\usepackage{amsmath}
\usepackage{float}
\usepackage{xcolor}
\DeclareFontSeriesDefault[rm]{bf}{m}

\shorttitle{EV Lac}
\shortauthors{Paudel et al.}

\begin{document}

\title{Simultaneous Multiwavelength Flare Observations of EV Lacertae}

\correspondingauthor{Rishi R. Paudel}
\email{rishi.paudel@nasa.gov}

\author[0000-0002-8090-3570]{Rishi R. Paudel}
\affiliation{University of Maryland, Baltimore County, Baltimore, MD 21250, USA}
\affiliation{CRESST II and Exoplanets and Stellar Astrophysics Laboratory, NASA/GSFC, Greenbelt, MD 20771, USA}

\author[0000-0001-7139-2724]{Thomas Barclay}
\affiliation{University of Maryland, Baltimore County, Baltimore, MD 21250, USA}
\affiliation{NASA Goddard Space Flight Center, Greenbelt, MD 20771, USA}

\author{Joshua E. Schlieder}
\author{Elisa V. Quintana}
\affiliation{NASA Goddard Space Flight Center, Greenbelt, MD 20771, USA}

\author[0000-0002-0388-8004]{Emily A. Gilbert}
\affiliation{Department of Astronomy and Astrophysics, University of Chicago, 5640 S. Ellis Ave, Chicago, IL 60637, USA}
\affiliation{University of Maryland, Baltimore County, 1000 Hilltop Circle, Baltimore, MD 21250, USA}
\affiliation{The Adler Planetarium, 1300 South Lakeshore Drive, Chicago, IL 60605, USA}
\affiliation{NASA Goddard Space Flight Center, Greenbelt, MD 20771, USA}

\author[0000-0002-5928-2685]{Laura D. Vega}
\affiliation{Department of Astronomy, University of Maryland, College Park, MD 20742, USA}
\affiliation{NASA Goddard Space Flight Center, Greenbelt, MD 20771, USA}
\affiliation{CRESST II and Exoplanets and Stellar Astrophysics Laboratory, NASA/GSFC, Greenbelt, MD 20771, USA}

\author{Allison Youngblood}
\affiliation{Laboratory for Atmospheric and Space Physics, 1234 Innovation Dr, Boulder, CO 80303, USA}

\author{Michele Silverstein}
\affiliation{NASA Goddard Space Flight Center, Greenbelt, MD 20771, USA}

\author{Rachel A. Osten}
\affiliation{Space Telescope Science Institute, 3700 San Martin Drive, Baltimore, MD 21218, USA}

\author[0000-0002-2471-8442]{Michael A. Tucker}
\altaffiliation{DOE CSGF Fellow}
\affiliation{Institute for Astronomy, University of Hawai'i, 2680 Woodlawn Drive, Honolulu, HI 96822, USA}

\author{Daniel Huber}
\author{Aaron Do}
\affiliation{Institute for Astronomy, University of Hawai'i, 2680 Woodlawn Drive, Honolulu, HI 96822, USA}

\author{Kenji Hamaguchi}
\affiliation{CRESST II and X-ray Astrophysics Laboratory NASA/GSFC, Greenbelt, MD, USA}
\affiliation{University of Maryland, Baltimore County, Baltimore, MD 21250, USA}

\author{D. J. Mullan}
\author{John E. Gizis}
\affiliation{Department of Physics and Astronomy, University of Delaware, Newark, DE 19716, USA}

\author{Teresa A. Monsue}
\author{Knicole D. Col\'{o}n}
\author{Patricia T. Boyd}
\affiliation{NASA Goddard Space Flight Center, Greenbelt, MD 20771, USA}

\author{James R. A. Davenport}
\affiliation{Department of Astronomy, University of Washington, Seattle, WA 98195, USA}

\author{Lucianne Walkowicz}
\affiliation{The Adler Planetarium, 1300 South Lakeshore Drive, Chicago, IL 60605, USA}

\DeclareFontSeriesDefault[rm]{bf}{m}
\begin{abstract}
We present the first results of our ongoing project conducting simultaneous multiwavelength observations of flares on nearby active M dwarfs. We acquired data of the nearby dM3.5e star EV Lac using 5 different observatories: NASA's Transiting Exoplanet Survey Satellite (TESS), NASA's Neil Gehrels Swift Observatory (\textit{Swift}), NASA's Neutron Interior Composition Explorer (NICER), the University of Hawaii 2.2-m telescope (UH88) and the Las Cumbres Observatory Global Telescope (LCOGT) Network. During the $\sim$25 days of TESS observations, we acquired three simultaneous UV/X-ray observations using \textit{Swift} that total $\sim$18 ks, 21 simultaneous epochs totaling  $\sim$98 ks of X-ray data using NICER, one observation ($\sim$ 3 hours) with UH88, and one observation ($\sim$ 3 hours) with LCOGT. We identified 56 flares in the TESS light curve with estimated energies in the range log $E_{\rm T}$ (erg) = (30.5 - 33.2), nine flares in the \textit{Swift} UVM2 light curve with estimated energies in the range log $E_{UV}$ (erg) = (29.3 - 31.1), 14 flares in the NICER light curve with estimated minimum energies in the range log $E_{N}$ (erg) = (30.5 - 32.3), and 1 flare in the LCOGT light curve with log $E_{L}$ (erg) = 31.6. We find that the flare frequency distributions (FFDs) of TESS and NICER flares have comparable slopes, $\beta_{T}$ = -0.67$\pm$0.09 and $\beta_{N}$ = -0.65$\pm$0.19, and the FFD of UVOT flares has a shallower slope ($\beta_{U}$ = -0.38$\pm$0.13). Furthermore, we do not find conclusive evidence for either the first ionization potential (FIP) or the inverse FIP effect during coronal flares on EV Lac.
\end{abstract}

\keywords{stars: flare--stars: individual: EV Lacertae }
\section{Introduction}
\label{sec:introduction}
M dwarfs, commonly known as red dwarfs, are the most abundant ($\sim$75\%) stars in our galaxy \citep{2006AJ....132.2360H}. They are low mass objects with masses \textbf{$\lesssim$ 0.6 $M_{\odot}$} \citep{2001RvMP...73..719B} and are considerably cooler and less luminous than the Sun. 
Due to their convective interiors and rotation, they have \textbf{relatively strong magnetic fields for their size} \citep{2017NatAs...1E.184S} and are capable of producing very strong flares with energies up to 10$^{4}$ times or greater than the strongest flare observed on the Sun (e.g., \citealt{2016ApJ...829...23D,2016ApJ...832..174O,2018ApJ...858...55P}). \textbf{It has been shown that M dwarfs of all ages are capable of producing flares} \citep{2014ApJ...781L..24S,1996AJ....112.2799H,2018ApJ...861...76P,2018ApJ...858...55P,2020AJ....160..237F}.

In the standard picture of a solar flare, energy release is governed by magnetic reconnection in the corona or upper chromosphere. During a flare, magnetic energy stored in magnetic fields is suddenly released in the form of kinetic energy of particles (ions and electrons), bulk plasma motion, and thermal emission mostly in the form of soft X-rays. Thermal coronal emission (soft X-rays) is produced as a result of heating by the non-thermal electrons produced in the corona. Those electrons also travel along field lines and emit gyrosynchrotron radio emission.  Electrons that are accelerated along field lines to the intersections of a magnetic loop with the photosphere produce bremsstrahlung seen in hard X-rays \citep{2016hasa.book.....S}. Flare blackbody (BB) emission is often seen in the UV and optical, sometimes correlating with the steep rise/impulsive phase as seen in X-ray observations \citep{2010ARA&A..48..241B}. This BB emission is a result of local heating in the chromosphere/photosphere by the particles which precipitate downwards after losing their energy in the form of hard X-rays. \cite{2013ApJS..207...15K} estimated the typical temperatures of the BB emission to be in the range $\sim$9,000 - 14,000 K: at such temperatures, the peak of the spectrum occurs in the near UV, at wavelengths of order 3000 \AA. The BB radiation escapes from the star in the form of a ``white light flare (WLF)".

The cumulative flare frequency distribution (FFD) of a flaring star has been found typically to follow a power-law. As a result, the FFD can be fit by a linear relation : log $\tilde{\nu}$ = C + $\beta$log $E$, where $\tilde{\nu}$ is the cumulative frequency defined as the number of flares per unit time  with energies in excess of $E$ (e.g. \citealt{2005stam.book.....G,1976ApJS...30...85L}). Each star is observed to have its own particular values of the coefficient $C$ and the spectral index $\beta$.

Since M dwarfs commonly host small planets on short period orbits, M dwarf planets may be exposed to extreme space weather environments and run the risk of being exposed to the enhanced electromagnetic radiation (mainly X-rays and UV radiation) and energetic particle flux coming from flares. This is particularly important when considering habitability because  M dwarf HZs are very close to the stars \citep{2013ApJ...765..131K}. For example, the planet Proxima Centauri b receives 30 times more extreme UV (10 – 121 nm) flux than Earth, 10 times more far UV (122 – 200 nm) flux, and 250 times more X-rays (0.01 - 10 nm, \citealt{2016A&A...596A.111R}). Although certain UV and optical photons from flares can have beneficial effects on life \citep{2017ApJ...843..110R,2018ApJ...865..101M}, negative effects of energetic photons and particles are likely to occur. High energy radiation may have adverse effects on the thermo-chemical equilibrium of the planets' atmospheres. This has many consequences including the loss of surface water, stripping of the planet's entire atmosphere or destruction of the ozone layer \citep{2007AsBio...7..185L,2010AsBio..10..751S,2017MNRAS.464.3728B,2019AsBio..19...64T,2020NatAs.tmp..248C}. To fully account for the impact of M dwarf flares on exoplanet atmospheres, we must constrain the total energy emitted during the flares on M dwarfs at various wavelengths.

The \textit{Transiting Exoplanet Survey Satellite mission} (TESS; \citealt{2014JAVSO..42..234R}) was launched in April 2018 to perform a near-all-sky photometric survey to find small planets around the brightest nearby stars, but also has sensitivity to low amplitude, short duration events, like flares. Its photometric bandpass ($\sim$600 - 1000 nm) is more sensitive at redder wavelengths compared to $Kepler$ \citep{2017PAPhS.161...38B}, and combined with its all-sky observing strategy, is ideal for targeting M dwarfs \citep{2015ESS.....350301R,2018ApJS..239....2B,2019AJ....157..113B}. 

Long-baseline, high-precision optical data-sets from TESS allow us to observe the diversity of flaring events with amplitudes spanning more than 5 orders of magnitude. However, atmospheric stripping of planets is caused by the strongly photo-dissasociative UV and X-ray photons, not optical photons \citep{2017MNRAS.464.3728B}. We cannot draw strong conclusions about habitability from optical data alone without first measuring the relationship between X-ray/UV and optical. Large surveys of the high energy radiation of M dwarfs such as the HST MUSCLES Treasury Survey \citep{2016ApJ...820...89F} and HAZMAT \citep{2014AJ....148...64S} have provided detailed UV flaring information on M dwarfs \citep{2018ApJ...867...70L,2018ApJ...867...71L}, but the link between optical and UV flares remains elusive.

In this paper we describe the first results from our large program studying nearby active flaring M dwarfs using multiwavelength datasets. We focus on the flaring M dwarf EV Lac, which has been known as a flare star for at least 65 years \citep{1955PASP...67...34R}. EV Lac produces flares in the X-ray (e.g., \citealt{1994ApJS...90..735S,1999A&A...342..502S,2000A&A...353..987F,2010ApJ...723.1558H}), UV (e.g. \citealt{1986ESASP.263..137A,1995AAS...186.2103P}), optical (e.g. \citealt{1976PASJ...28..665K,1995A&AS..114..509A}) and radio wavelengths (e.g. \citealt{1989ApJS...71..895W,1995A&AS..114..509A}).

\cite{2005ApJ...621..398O} \textbf{carried out a simultaneous multiwavelength observing campaign of EV Lac} for two days in 2001 September using radio (VLA), optical (McDonald Observatory), UV ($HST$), and X-ray ($Chandra$) telescopes. They observed a large flare at radio wavelengths, two small flares at both optical and UV wavelengths, and at least nine flares in X-ray. A very large flare occurred on this star in 2008 April,  which resulted in a trigger from {\it Swift}'s autonomous gamma-ray burst response \citep{2010ApJ...721..785O} and is one of the most extreme stellar flares observed in terms of its enhancement relative to the quiescent level. Its peak flux of 5.3 $\times$ 10$^{-8}$ erg cm$^{-2}$ s$^{-1}$  at 0.3 - 100 keV was estimated to be $\sim$ 7000 times the star's quiescent X-ray flux and in white-light the star brightened by $\geq$ 4.7 mag. At the flare peak, it had L$_{X}$/L$_{\rm bol}$ $\sim$ 3.1, where $L_{\rm bol}$ is the bolometric luminosity of the star during the early stages of the flare. \\ %
\indent Since EV Lac is nearby and known to produce flares frequently across the electromagnetic spectrum, it is one of the best targets for simultaneous multi-wavelength observations.

We observed EV Lac using three space telescopes: TESS, \textit{Swift}, and NICER and two ground based telescopes: The University of Hawaii 2.2-meter telescope (UH88) and a one meter telescope at McDonald Observatory as part of the Las Cumbres Observatory Global Telescope Network (LCOGT, \citealt{2011IAUS..276..553S}). In Section \ref{sec:target_characteristics}, we give a brief introduction to EV Lac and in Section \ref{sec:data_sets} we describe the various observations. Likewise, in Section \ref{sec:analysis}, we present the data reduction and flare analysis. In Section \ref{sec:summary_discussion}, we discuss and summarize the main results of our work. 
\section{Target Characteristics} \label{sec:target_characteristics}
\begin{table}
 	\caption{Properties of EV Lac}
 	\label{table:properties}
     \centering
     \scalebox{0.6}{
     \begin{tabular}{cccc}
     \hline
       & Value & Units & Ref. \\
       \hline
       ASTROMETRIC PROPERTIES \\
       \hline
        $\alpha$ & 341.7029626 ($\pm$0.03 mas) & deg & 4 \\
        $\delta$ & +44.3320170 ($\pm$0.03 mas) & deg & 4 \\
        $\mu_{\alpha}$ & -706.1$\pm$0.1 & mas yr$^{-1}$ & 4 \\
        $\mu_{\delta}$ & -458.8$\pm$0.1 & mas yr$^{-1}$ & 4 \\
        parallax & 198.01$\pm$0.04 & mas & 4 \\
        distance & 5.049$\pm$0.001 & pc & 15 \\
       \hline
        PHOTOMETRIC PROPERTIES \\
       \hline
      $V_J$    & 10.22$\pm$0.03 & mag & 11 \\
      $R_{KC}$ &  9.05$\pm$0.03 & mag & 11 \\
      $I_{KC}$ &  7.55$\pm$0.02 & mag & 11 \\
      $J$ & 6.11$\pm$0.03 & mag & 2 \\
      $H$ & 5.55$\pm$0.03  & mag & 2 \\
      $K_{s}$ & 5.30$\pm$0.02 & mag & 2 \\
      $i$ & 13.215$\pm$0.002 & mag & 3 \\
      $G$ & 9.000$\pm$0.001 & mag & 4 \\
      $BP$ & 10.543$\pm$0.004 & mag & 4 \\
      $RP$ & 7.809$\pm$0.001 & mag & 4 \\
      $T_{mag}$ & 7.73$\pm$0.01 & mag & 5 \\
      $W1^b$ & 5.241$\pm$0.063 & mag & 12, 13 \\
      $W2^b$ & 4.643$\pm$0.042 & mag & 12, 13 \\
      $W3^b$ & 4.891$\pm$0.015 & mag & 12, 13 \\
      \hline
      PHYSICAL PARAMETERS\\
       \hline
        Sp. Type & dM3.5e  &  & 1 \\
        T$_{eff}$ & 3270$\pm$80 & K & ** \\
        $M$ & 0.347$\pm$0.020 & $M_{\odot}$  & ** \\
        $R$ & 0.353$\pm$0.017 & $R_{\odot}$ & ** \\ 
        $L_{bol}$ & 0.0128$\pm$0.0003 & $L_{\odot}$ & ** \\ 
        log {\it g} & 4.89$\pm$0.00 & $\log$(cm s$^{-2}$) & 5 \\
        p$_{rot}$ & 4.38 & d & 8 \\
        \hline
        SPECTRAL PROPERTIES \\
        \hline
        [Fe/H] & -0.01 $\pm$ 0.15 & dex & 14 \\
        RV   & 0.19 & km s$^{-1}$ & 7 \\
        $v$ sin$i$ & 3.50 & km s$^{-1}$ & 7 \\
             \hline
             ACTIVITY INDICATORS \\
             \hline
             EW H$\alpha$ & -4.54$\pm$0.04 & \AA & 9 \\
             EW Ca II K & 14.86 & \AA & 10 \\
             log $L_{H \alpha}$/$L_{bol}$ & -3.76 & & 9 \\
             log $L_{X}$/$L_{bol}$ & -3.33 & & 6 \\
             log $R'_{HK}$ & -4.24$\pm$0.11 & & 16 \\
             \hline
             \end{tabular}}
              \\
              {\textbf{Note}: $^{a}$epoch J2015.5\\
                $^{b}$All-Sky Data Release\\
                $^{**}$This work. \\
               \textbf{References:}} \\
              (1) \cite{1995AJ....110.1838R}; (2) \cite{2003tmc..book.....C} \\
              (3) \cite{2016arXiv161205560C};
             (4) \cite{2018arXiv180409365G}; \\(5) \cite{2018AJ....156..102S};
             (6) \cite{2008MNRAS.390..567M};\\
             (7) \cite{2018A&A...612A..49R}; (8) \cite{1980AJ.....85..871P};\\ (9) \cite{2017ApJ...834...85N};
             (10) \cite{2017ApJ...843...31Y}; 
             (11) \cite{Weis1996};
             (12) \cite{Wright2010};
             (13) \cite{Cutri2012};
             (14) \cite{RojasAyala2012};
             (15) \cite{2018AJ....156...58B};
             (16) \cite{2020AJ....160..269M}
\end{table}
EV Lac (GJ 873, LHS 3853), at a distance of only 5.05 pc \citep{2018arXiv180409365G}, is one
of the most widely studied low-mass stars. In order to measure accurate flare energies, we require a self-consistent set of stellar parameters. We use two methods to estimate the star's 
fundamental properties that include its highly precise \textit{Gaia} parallax. We used the 
metallicity dependent M$_{Ks}$–radius relationship of \citet{Mann2015} to estimate
the stellar radius, adopting [Fe/H] = -0.01$\pm$0.15 as determined by \cite{RojasAyala2012}. To calculate the star's effective temperature (T$_{\rm eff}$),
we used the relations of \citet{Mann2015} to estimate the $K$-band bolometric correction, 
calculated the luminosity, and then substituted luminosity and radius
into the Stefan–Boltzmann law. 
Fundamental parameter uncertainties were estimated via 
Monte Carlo methods where we adopted Gaussian-distributed measurement errors and added the 
systematic scatter in each parameter. We estimate $R_{*}$ = 
0.337 $\pm$ 0.029 $R_{\odot}$, L$_{*}$ = 0.0124 $\pm$ 0.0007 $L_{\odot}$, and $T_{\rm eff}$ = 
3315 $\pm$ 152 K. However, in view of the rather large error bars, we consider that it is possible to arrive at more precise values of the parameters of EV Lac by using a different approach as follows.

For comparison, we now turn to the methods of Silverstein et al.~(2021, in preparation), which are heavily based on those of \cite{Dieterich2014}. To derive effective temperature, we assume [Fe/H] = 0 and compare an assortment of color combinations from observed \cite{Weis1996} $V_JR_{KC}I_{KC}$, 2MASS $JHK_S$ \citep{2003tmc..book.....C}, and WISE All-Sky $W1W2W3$ \citep{Wright2010, Cutri2012} photometry to those extracted from scaled BT-Settl 2011 CIFIST model spectra \citep{Allard2012}. We note that W2 appears brighter by 0.25 mag than W3, and both W1 and W2 bands are affected by saturation; we suspect the excess in W2 is saturation-induced, rather than from an infrared source such as a circumstellar disk. We believe \textbf{the impact of potentially underestimated W1 and W2 magnitudes is} likely negligible in our analysis because these bands are far into the Rayleigh-Jeans tail. \textbf{Repeating the procedure 16 times, with permutations of W1 and/or W2 values varied by $+0.1$, $+0.2$, or $+0.3$ magnitudes, yielded effective temperature values ranging from 3270 K to 3340 K, equal to or within the error bars of our adopted value of 3270 $\pm$ 80 K.  Once an effective temperature is derived, we iteratively modify the model spectrum closest to our results using a polynomial scaling factor until the observed and new model photometry match to within 0.063 mag. This 0.063 value corresponds to the largest error bar in our observed photometry.} We integrate the resulting spectrum across the wavelength range of our observed photometry and apply a bolometric correction based on the amount of flux expected in the remaining wavelengths of a blackbody of the same temperature. We scale the final flux by the \textit{Gaia} DR2 parallax \citep{2018arXiv180409365G} to calculate bolometric luminosity and derive a radius using the Stefan-Boltzmann Law. \textbf{Varying W1 and W2 as described earlier yields only a small range of luminosity values (0.0125 $L_\odot$ - 0.0130 $L_\odot$) and radius values (0.340 $R_\odot$ - 0.353 $R_\odot$), on the order of our error bars.
We estimate 
R$_*$ = 0.353 $\pm$ 0.017 R$_{\odot}$, 
L$_{*}$ = 0.0128 $\pm$ 0.0003 L$_{\odot}$ (log$_{10}$L$_{*}$ = -1.894 $\pm$ 0.011)}, and 
T$_{\rm eff}$ = 3270 $\pm$ 80 K. We also estimate the mass of EV Lac using the M$_K$ - mass relation of \cite{2016AJ....152..141B} and find 0.347 $\pm$ 0.020 M$_{\odot}$. These stellar parameters are consistent with those estimated using the \citet{Mann2015} relations. We adopt these parameters for EV Lac and list them in Table \ref{table:properties}. We also use the scaled model spectrum derived using the methods detailed here in our white light flare analysis presented in \S~\ref{sec:analysis}.


We compile these properties, additional properties, and their literature references in Table~\ref{table:properties}. The star's mass, radius, and spectral type of dM3.5e place it close to, but later than, the range of spectral types (dM2e-dM3e) where main sequence stars are believed to make a transition between partially convective and fully convective interiors \citep{2017ApJ...837...96H}. This fully convective structure in EV Lac along with its rotation period of \textbf{$\sim$4.4} days results 
in strong magnetic activity. Previous observations provided constraints on the star's magnetic fields, revealing they cover $>$50\% of the stellar surface and have strengths of $\approx$4 kG  \citep{1996ApJ...459L..95J,1994svsp.coll..147S}. This magnetic 
activity manifests itself in the form of star spots, flares, and associated high energy emission. Observationally, EV Lac is found to be the second brightest X-ray source seen in the $ROSAT$ All-Sky Survey \citep{1999A&AS..135..319H}.
%

\subsection{Stellar Age}

Aspects of EV Lac's high level of magnetic activity may also be traced to its age. Here we investigate multiple properties of the star that in aggregate provide an age constraint in order to study the star's flare properties in the context of other targets in the ``M Dwarf Flares Through Time" program \footnote{TESS Guest Investigator programs G011266 and G022252, PI J. Schlieder, and G03226, PI M. Silverstein}.

\textit{HR Diagram Position} - The slow evolution of low-mass stars like EV Lac as they contract to the main sequence provides the means to estimate ages via position on the Hertzsprung-Russel Diagram (HRD). Assuming similar metallicities, younger M dwarfs appear brighter than older stars. The advent of precision \textit{Gaia} parallaxes and photometry allows for comparisons of EV Lac's HRD position to similar stars in populations with well determined ages. We use EV Lac's \textit{Gaia} parallax and photometry \citep{2018arXiv180409365G} to calculate its absolute magnitude ($M_G = 10.484$) and color ($BP-RP = 2.735$) in the \textit{Gaia} bands, and compare its HRD position to populations of low-mass stars of known age presented in \cite{Gaia_HRD2018}. \textbf{EV Lac is  $\approx$0.4 mag fainter} than the majority of similar color stars in the $\sim$110-125 Myr old Pleiades cluster \citep{Stauffer1998, Dahm2015}, and  \textbf{has a similar absolute magnitude} when compared to low mass stars in the $\sim$600-800 Myr old Praesepe \citep{Brandt2015a, Douglas2017} and Hyades \citep{Brandt2015b, Douglas2019} clusters. These comparisons suggest an age of $>$125 Myr for EV Lac, but do not provide a stringent limit. 


\textit{X-ray Emission} - We use EV Lac's measured ROSAT count rate and hardness ratios from the 2RXS catalog of \cite{Boller2016} and the count rate to flux conversion from \cite{Schmitt1995} to calculate an X-ray flux of 4.13 x 10$^{-11}$ ergs cm$^{-2}$ s$^{-1}$. We then combined this with the \textit{Gaia} DR2 stellar distance to estimate an X-ray luminosity of 1.27 x 10$^{29}$ erg s$^{-1}$.   We compared this luminosity with the X-ray properties of the young to old populations presented in \cite{Bowler2012}. \textbf{Comparing to \cite{Bowler2012}, their Figure 5, EV Lac has an X-ray luminosity that matches the Pleiades cluster distribution within 1$\sigma$ of the median and matches the Hydaes cluster distribution within 2$\sigma$ of the median. In the same \cite{Bowler2012} figure, EV Lac's X-ray luminosity is more than 250$\times$ larger than the median luminosity of old low-mass stars in the galactic field.} 

\textit{Rotation} - EV Lac is a relatively rapid rotator, with P$_{rot}$ = 4.38 days measured via periodic brightness modulations in SUPERWASP photometry \citep{2006PASP..118.1407P}. This rotation period is consistent with the period we estimate using the TESS photometry (see \S~\ref{sec:analysis}). In the period-color diagrams presented in \cite{Curtis2020}, this rotation period places EV Lac among other M dwarfs in Praesepe, but also close to the slowest rotators in the Pleiades. As an additional constraint, we also use EV Lac's rotation period and the M dwarf rotation-age relation of \cite{Engle2018} to estimate an age of 280$^{+220}_{-230}$ Myr, consistent with the general range of ages inferred from other diagnostics.  

\textit{Kinematics} - 
        Using probabilistic methods to study membership in stellar kinematic groups, \cite{2014A&A...567A..52K} suggest EV Lac may be a member of the Ursa Major moving group \citep[UMaG,][]{Proctor1869, Roman1949, King2003}, which would indicate an age of $\approx$400 Myr \citep{2015ApJ...813...58J}. However, \cite{2012ApJ...758...56S} do not associate EV Lac with any of the moving groups they study, including UMaG. We use the star's updated \textit{Gaia} astrometry to calculate its galactic velocities and compare to the revised Ursa Major group properties presented in \cite{2018ApJ...856...23G}.  Following the methods of \cite{1987AJ.....93..864J}, we calculate ($UVW$)$_{EV Lac}$  = (+19.765, +3.596, -1.709) $\pm$ (0.004, 0.002, 0.002) km s$^{-1}$. For the UMaG, \cite{2018ApJ...856...23G} report ($UVW$)$_{UMaG}$ = (+14.8, +1.8, -10.2) km s$^{-1}$. EV Lac's galactic velocities are broadly consistent with the UMaG in $U$ and $V$, but it is a significant outlier in $W$. The star's galactic position is also removed from the core of the UMaG group, lying $\sim$20 pc away from the tightly clustered nucleus described by \cite{2010AAS...21545505M} and \cite{2016ApJ...818....1S}. We do note that EV Lac's kinematics place the star among the proposed UMaG stream members proposed by \cite{King2003}, but the membership of many of these stars remains unconfirmed.     
As a final check, we use BANYAN $\Sigma$ \citep{2018ApJ...856...23G}, a Bayesian analysis tool which estimates the probability of kinematic group membership. BANYAN $\Sigma$ suggests that EV Lac is not a kinematic member of the UMaG (0\% probability) or any other group included in the analysis. The star's kinematics are broadly consistent with other young stars in the solar neighborhood, but group membership cannot be confirmed.

\textit{Age Summary} - In aggregate, the available observations and calibrated samples for comparison indicate that EV Lac is an intermediate age M dwarf. Its HR Diagram position indicates it is likely older than the $\sim$125 Myr Pleiades cluster while its X-ray luminosity and rotation rate suggest an age comparable to the 600 - 800 Myr Hyades and Praesepe clusters and perhaps younger. The star's galactic velocities and positions are also broadly consistent with proposed members of the $\sim$400 Myr UMaG kinematic stream, but its membership remains inconclusive. Given these properties, we quantitatively place EV Lac in the 125 - 800 Myr age range.

%


\section{Data Sets}
\label{sec:data_sets}
\subsection{TESS}
TESS observed EV Lac (TIC 154101678, GJ 873, 2MASS J22464980+4420030) during Sector 16 (11 Sep, 2019 - 07 Oct, 2019) as a part of its Cycle 2 observations. EV Lac was observed in two-minute cadence as a part of proposals G022252, G022198, G022080, and G022056. The total TESS observation time is 23.2 days. We used the SAP flux for our analysis after filtering using the `hard' bitmask option in the data analysis tool \textbf{`Lightkurve' \citep*{lightkurve}} and removing NaNs from the data.
\subsection{Swift/{\normalfont XRT} Data}
EV Lac was observed by \textit{Swift}'s X-ray telescope (XRT; \citealt{2005SSRv..120..165B}) three times on 2019 September 21-22, via the mission's Target of Opportunity (ToO) program (\#12734 and \#12758). The XRT is mainly designed to observe soft X-rays in the energy range of 0.3-10 keV using CCD detectors and has an energy resolution of $\approx$ 140 eV in the region of the Fe K-line at E=6.4 keV. The first observation occurred on 2019 September 21 at UT 12:34:57 for a time interval of 7.1 ks, the second on 2019 September 22 at UT 11:02:32 for 8.0 ks, and the third on 2019 September 23 at UT 13:51:18 for 2.9 ks. The observing IDs for the three observations are 00031397002, 00031397003, and 00031397004 respectively. 

The star was observed in Photon Counting (PC) mode, as well as Windowed Timing (WT) mode for very short intervals. WT mode is preferred whenever the count rate is very high, thereby causing saturation in CCD detectors in the PC mode. Since no big events occurred on the star during the Swift observations, we analyzed the data collected in PC mode only. We obtained the raw data from UK Swift Science Data Centre (UKSSDC). We used \textit{Swift} \texttt{XRTPIPELINE} task (version 0.13.5) and calibration files from the High Energy Astrophysics Science Archive Research Center (HEASARC)'s calibration database system (CALDB; index version = `x20190910') to reduce the raw data and produce cleaned and calibrated files. We obtained the X-ray light curve using HEASARC's \texttt{XSELECT}, a high level command interface to the HEASARC FTOOLS, to extract a circular region of radius 30 pixels centered at the position of the source (RA = 341.70$^{o}$, DEC = +44.33$^{o}$). \textbf{A 30 pixel radius circle encloses $\sim$95\% of the PSF for a bright source} \footnote{https://www.swift.ac.uk/analysis/xrt/files/xrt\_swguide\_v1\_2.pdf}.  \textbf{We found that the average photon count rate is 0.40 counts s$^{-1}$ for the quiescent level\footnote{The enhanced X-ray events were excluded while estimating the quiescent level.}}, so the pile-up correction was not applied. The pile-up generally occurs when the count rate is high ($\gtrsim$ 0.5 counts $^{-1}$)\footnote{\url{https://www.swift.ac.uk/analysis/xrt/pileup.php}} so that multiple photons registered in a given CCD detector have overlapping charge distributions. This may result in an incorrect classification of a true X-ray event. We used only grade 0-12 events in the PC mode, which are considered to be good for science.\\
\indent We used the \texttt{BARYCORR}\footnote{\url{https://heasarc.gsfc.nasa.gov/ftools/caldb/help/barycorr.html}} FTOOL to perform barycenter correction on our XRT data, and \texttt{XSELECT} to extract source and background spectra from the cleaned event list. For this, the same circular extraction region described above was used for the source. For the background, we used an annular extraction region with inner radius of 40 pixels and outer radius of 70 pixels centered at the source position. The exposure maps were prepared while running XRTPIPELINE using option \texttt{createexpomap=yes}, and the ancillary response file (ARF) was produced using \texttt{XRTMKARF} which needs an XRT response matrix file (RMF). We used v014 RMF: swxpc0to12s6\_20130101v014.RMF obtained from the CALDB file.

\subsection{Swift/{\normalfont UVOT} Data}
The \textit{Swift} Ultra-Violet/Optical Telescope (UVOT; \citealt{2005SSRv..120...95R}) also observed EV Lac during the same time as the XRT. The first observation started on
21 September at UT 12:34:55, the second on 22 September at UT 11:02:29, and the third on 23 September at UT 13:51:20. All three observations were performed with the UVM2 filter centered at $\lambda$ = 2259.84 \AA ($\lambda_{min}$ = 1699.08 \AA, $\lambda_{max}$ = 2964.30 \AA, FWHM = 527.13 \AA). The raw data were obtained from UDSSDC, and then processed in two steps to obtain a cleaned event list. First, we used \texttt{COORDINATOR}\footnote{\url{https://heasarc.nasa.gov/ftools/caldb/help/coordinator.html}} to convert raw coordinates to detector and sky coordinates. Second, we used \texttt{UVOTSCREEN} to filter the hot pixels and obtain a cleaned event list. \\
\indent A calibrated light curve was extracted from the cleaned event list by using the FTOOL \texttt{UVOTEVTLC}. For the source, \textbf{we used the recommended circular extraction region of radius of 5$\arcsec$ around the source position}, and for a smooth background, a circular extraction region of radius 30$\arcsec$ away from source. Furthermore, we used \texttt{timebinalg = u} to bin time by 11.033 s. It is required that the bin size be a multiple of the minimum time resolution of UVOT data which is 11.033 ms. \texttt{UVOTEVTLC} applies a coincidence loss correction whenever there is pile up of photons on detectors, by using the necessary parameters from CALDB. After this, the light curve was  barycenter corrected by using \texttt{BARYCORR}. To place all the observatories on a common time system, the barycentric times were then converted to the Modified Julian Date (MJD) system. 

%
%
\subsection{Neutron Star Interior Composition Explorer ({\normalfont NICER})}
\label{subsection:NICER data}
During TESS Sector 16, EV Lac was observed simultaneously by NASA's NICER X-ray mission \citep{2016SPIE.9905E..1HG}, via the Target of Opportunity (ToO) program.  NICER was designed to study soft X-rays within the energy band 0.2-12 keV, with high signal-to-noise-ratio photon counting capability. It has an X-ray Timing Instrument (XTI) with time-tagging resolution of $<$ 300 nsec (absolute), which is much better than other current X-ray missions (e.g., 100-1000 times better than XMM). Its energy resolution is similar to those of the XMM and Chandra non-grating CCD instruments (137 eV at 6 keV).

NICER observed EV Lac on 21 different days for a total exposure time of $\sim$98 ks (Observation IDs: 21004201[25-45]). We obtained calibrated and cleaned event files of our observation from the NICER archive\footnote{\url{https://heasarc.gsfc.nasa.gov/docs/nicer/nicer_archive.html}}. The cleaned event files were obtained from raw data using the NICER-specific HEASoft tool \texttt{NICERDAS}\footnote{\url{https://heasarc.gsfc.nasa.gov/docs/nicer/data_analysis/nicer_analysis_guide.html}}. They were barycenter corrected by using \texttt{BARYCORR}. We then used \texttt{XSELECT} to generate light curves. For the spectral analysis, we applied the latest calibration (ver. 20200722) to the unfiltered event data and processed the data through the standard screening criteria using the \texttt{nicerl2} FTOOL. NICER does not provide spatial information of the source, but it does provide timing and energy information of each photon. So it is not possible to extract neither light curves nor spectra by using source and background extraction regions. However, the mission provides background estimator tools to estimate the background spectra. We used the \texttt{nibackgen3C50} (v6) tool for extracting source spectra and estimating background spectra of the corresponding time intervals. This tool uses a background events file which was created from ``blank sky" observations by NICER. The ``blank sky" region was based on the Rossi X-ray Timing Explorer (RXTE) background fields. \textbf{Detectors 14 and 34 are known to suffer from increased noise.} So we excluded data from those detectors and used the g2020a background model to generate background spectra. We produced detector response (rmf and arf) files for these spectra,
by following the instruction at section “Calculating ARF and RMF for Different Subset of Modules” in \url{https://heasarc.gsfc.nasa.gov/docs/nicer/analysis_threads/arf-rmf/}. 
\subsection{University of Hawaii 2.2 meter ({\normalfont UH88}) Telescope}
We were awarded two nights to observe EV Lac with the SNIFS instrument \citep{2004SPIE.5249..146L} on the University of Hawaii 2.2 meter telescope (UH88) at Mauna Kea Observatories, 2019 September 20 and 2019 September 21. SNIFS has two modules, a blue arm and a red arm, which combine to cover a broad wavelength range. The blue arm covers 320 - 560 nm with a resolving power $\sim$1000 at 430 nm.  The red arm covers and 520 - 1000 nm with a resolving power $\sim$1300 at 760 nm.

On September 20, the humidity was above allowable limits and we were unable to open. On September 21, we obtained 10 spectra of EV Lac with 90 second exposures and 11 with 30 second exposures. These spectra showed significant variability due to variable cloud coverage throughout the night and the last few exposures were completely contaminated by clouds.

Due to the limited amount of time with no clouds and low enough humidity, we only obtained $\sim$ 3 hours worth of monitoring data of EV Lac. Therefore, the time series spectra were not useful for flare monitoring. During the period we were observing, there was at least one small ($<$.5 \%) white light flare visible in the TESS photometric light curve. However, due to cloud variability, we were unable to discern any visible evidence of this flare in the spectra.

\begin{figure*}
    \centering
    \includegraphics[width=0.8\textwidth]{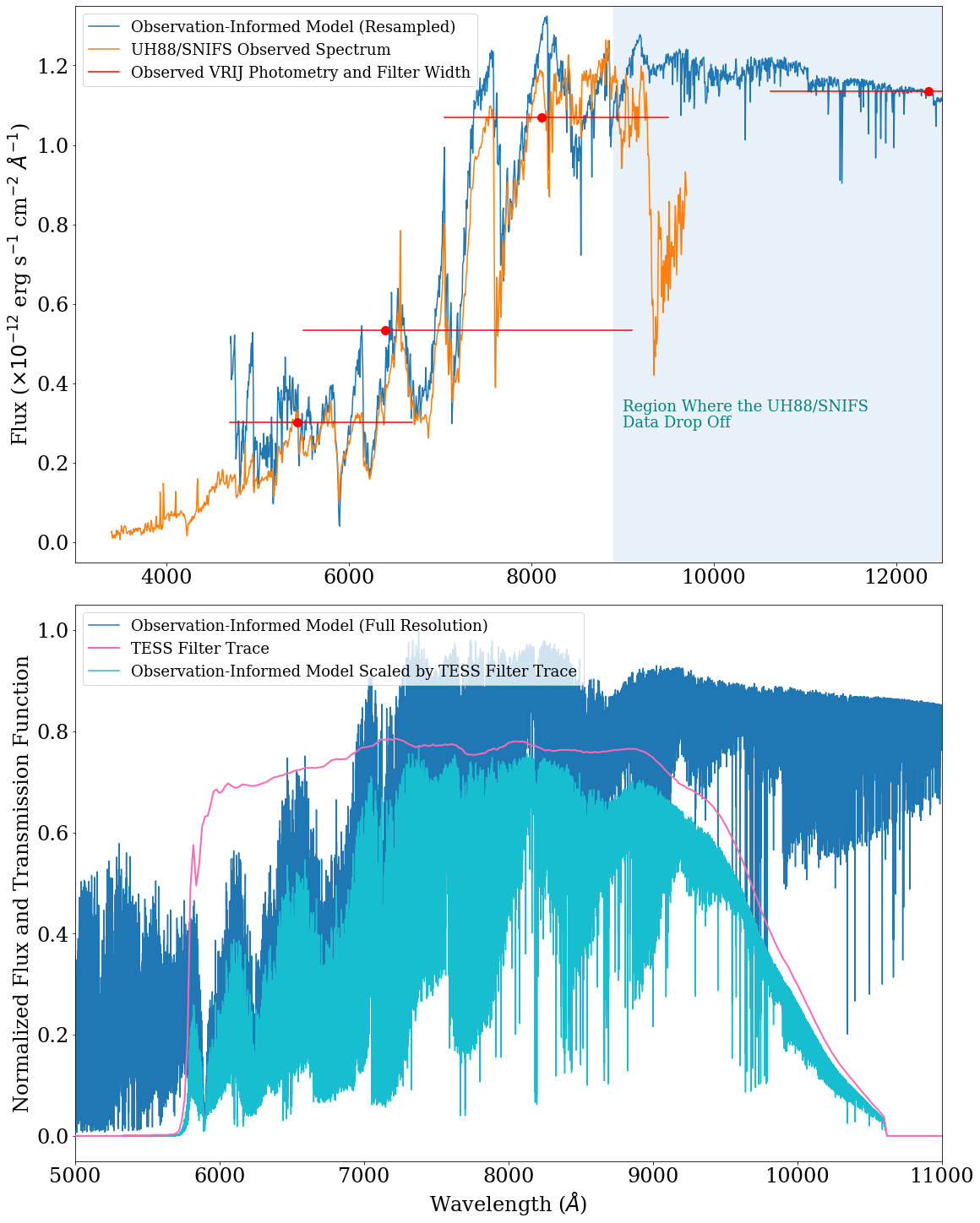}
    \caption{\textit{Top Panel}: The flux calibrated spectrum of EV Lac taken with UH88/SNIFS (orange) shows emission and absorption lines consistent with an active M dwarf. The observation-informed model spectrum (blue)  described in $\S$~\ref{sec:analysis} resampled to the UH88/SNIFS resolution and the photometry used to derive it (in that wavelength window, red) are overlaid to show the match between both spectra and the photometry. The points and lines correspond to the effective wavelength and width of each filter, respectively. The shaded region demonstrates where the UH88/SNIFS data drop off in quality and then end completely. Because the model spans the full wavelength range of the TESS filter and matches observations, we adopt it for our analysis in $\S$~\ref{sec:analysis}. \textit{Bottom Panel}: We multiplied the observation-informed model spectrum (blue) described in $\S$~\ref{sec:analysis} with the TESS transmission function (pink) in order to determine the total energy emitted by EV Lac within the TESS bandpass.}
    \label{fig:tess_model_UH88}
\end{figure*}

The spectra were reduced using the SNIFS reduction pipeline \citep{2001MNRAS.326...23B} and flux calibrated using archival photometry. We used the \cite{2013A&A...549A...8B} model to correct for atmospheric attenuation. We then confirmed the flux calibration by using \textit{Gaia} data \citep{gaia2013, gaia2018}. We examined one of our spectra with minimal impact from cloud cover (see Fig. \ref{fig:tess_model_UH88}) and confirm it is consistent with the previously estimated M3.5 spectral type of EV Lac. The spectrum also exhibits emission in several activity diagnostic lines, like H$\alpha$, as previously observed.

\subsection{Las Cumbres Observatory ({\normalfont LCOGT})} 
\begin{figure*}[htp]
   \centering
   \includegraphics[width=\textwidth]{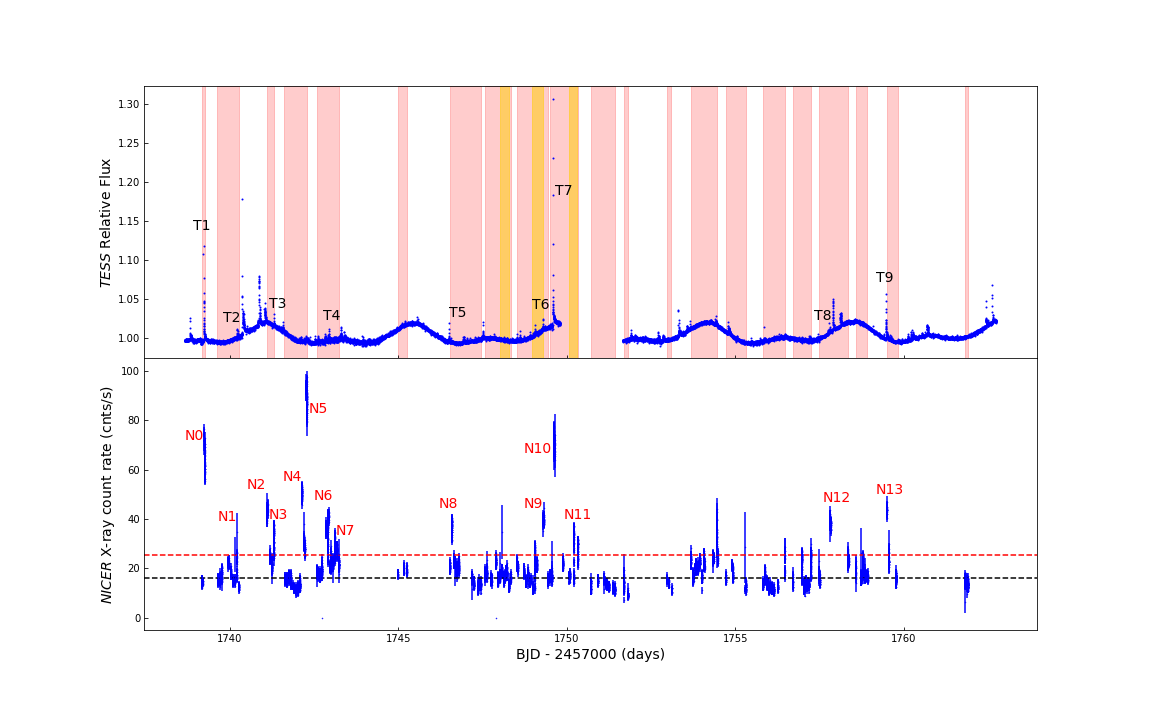}
   \caption{\textit{Upper Panel}: The EV Lac light curve obtained by TESS during Sector 16. Pink shaded regions correspond to times when the star was observed by NICER and the orange shaded regions correspond to times when it was observed by \textit{Swift}. The IDs: T1 through T9 represent the flares observed simultaneously by TESS and other facilities. These flare IDs are also listed in Table \ref{table:TESS-flares}. \textit{Lower Panel}: NICER X-ray light curve of EV Lac. We identify 14 flares in this light curve (labeled N0 through N13), but none were observed for their full duration. Some events have single point brightening or are due to background signal. The black dashed horizontal line corresponds to the quiescent level ($M_{N}$), and the red dashed line corresponds to $M_{N}$+2.5$\sigma$ value of count rate which we also consider as a threshold for identifying flares in this light curve.}
\label{fig:tess_obs_times}
\end{figure*}

EV Lac was observed by the LCOGT 1-m network on 2019 September 17, 2019 September 21, and 2019 September 23 as part of program NOAO2019B-001. We used 30 second exposures in the Bessell-U filter and obtained useful data over a period of $\sim$3 hours. 2019 September 21 was limited by weather, and 1.5 hours of useful photometry were collected each on 2019 September 17 and 2019 September 23. We used LCOGT's reduced images from the automatic pipeline software \texttt{BANZAI}, which performs bad-pixel masking, bias subtraction, dark subtraction, flat field correction, and applies an astrometric solution \citep{banzai}. We extracted aperture photometry of EV Lac and 8 comparison stars using \texttt{Photutils}, an Astropy package for
detection and photometry of astronomical sources \citep{2016ascl.soft09011B}. We observed one distinct flare and the slow decay of at least one other flare.\\ 
\indent In Table \ref{table:summary of obs. times}, we list the dates when EV Lac was observed by a given facility and the corresponding total observation time. The observation times of various facilities are also highlighted in the upper plot of Figure \ref{fig:tess_obs_times}. Each flare observed simultaneously by TESS and other facilities are given TESS IDs which are shown above each flare in the figure except for the flare T7. T5 is a small amplitude flare which cannot be seen in this figure. \\ 
\begin{table}[!htbp]
 	\caption{Summary of Observation Times}
 	\centering
     \begin{tabular}{ccc}
     \hline
      Facility & Date of obs. & time observed \\
       \hline
       TESS & 11 Sep - 07 Oct, 2019 & $\sim$25 d \\
       \textit{Swift} XRT & 21 - 23 Sep, 2019 & 18 ks \\
       \textit{Swift} UVOT & 21 - 23 Sep, 2019 & 18 ks \\
       NICER & 12 Sep - 05 Oct, 2019 & 97.7 ks \\
       UH88 & 20 - 21 Sep, 2019 & $\sim$3 hr \\
       LCOGT & 17 - 23 Sep, 2019 & $\sim$3 hr \\
       \hline
      \end{tabular}
\end{table}
\label{table:summary of obs. times}
%
%
From here onward, unless otherwise mentioned, the times are expressed in terms of TESS time which is BJD - 2457000 (days) for all the facilities.  
\section{Analysis}
\label{sec:analysis}
\subsection{White light flares observed by {\normalfont TESS}}

We conducted a white light flare analysis for EV Lac using one sector of TESS two-minute cadence data. Following the methods of \cite{pitkin14}, we pulled the light curve from the Mikulski Archive for Space Telescopes (MAST) using \texttt{lightkurve}. We then used an adapted version of \texttt{bayesflare} on the TESS data to detect stellar flares using Bayesian inference. The routine \texttt{bayesflare} uses a sliding window to inspect all of the data points by comparing them to a flare template and determining the odds that the data are best described by a flare with noise or just noise. Using this method, we identified 56 flares in the TESS light curve.

This flare detection routine returns the basic flare parameters we require to model the flares. We also used a Lomb-Scargle periodogram (LSP) to estimate a rotation period, see Figure \ref{fig:periodogram}. We then simultaneously model the smoothly varying light curve modulations caused by star spots and flares simultaneously. We built the model using a framework of \texttt{PyMC3} \citep{exoplanet:pymc3}, \texttt{celerite} \citep{celerite1,celerite2}, and \texttt{xoflares} \citep{xoflares}. We use a periodic Gaussian Processes (GP) to model the star spot modulated stellar rotation using a ``Rotation term" from \texttt{celerite}, and long term variability (with a jitter term to capture white noise) in the light curve. The LSP rotation period is used as a prior in this model. At the same time, we use the flare properties from \texttt{bayesflare} - full-width at half-maximum (FWHM), peak time, and peak amplitude - to seed the flare model using and sample the flare properties using \texttt{xoflares} \citep{xoflares}. We sampled over the posterior of our flare model to map the posterior distribution. We did this using \texttt{PyMC3}'s Automatic Differentiation Variational Inference algorithm (\textbf{ADVI; \citealt{2016arXiv160300788K}}) with 100,000 iterations and drawing 3,000 samples from the posterior distribution. We included as a model parameter the integral of each flare, which allowed us to determine posteriors on the flare energies.

In Figure \ref{fig:tess lc}, we show the TESS light curve in the top panel with the GP fit to the rotational modulation in green. We detrend the GP model from the data and show the flare model overlaid in pink in the middle panel. In the bottom panel we plot the residuals obtained after subtracting both the GP and flare models from the light curve. The flare parameters obtained after GP modeling are listed in Table \ref{table:TESS-flares}. The first column is the flare ID, the second is the flare peak time, the third is the FWHM, the fourth is the amplitude, the fifth column is the equivalent duration (ED), and the sixth column is the flare energy.

We used the modified model spectrum resulting from our spectral energy distribution analysis described in $\S$~\ref{sec:target_characteristics} to determine precise flare energies, as the UH88 spectrum did not cover the full TESS bandpass (Fig. \ref{fig:tess_model_UH88} top panel). The model matches both the UH88 spectrum and VRIJ photometry, which probe portions of the wavelengths covered by the TESS bandpass. We multiplied the model spectrum by the TESS transmission function to determine the energy emitted by EV Lac within the TESS bandpass (Fig. \ref{fig:tess_model_UH88} bottom panel). We then integrated across the whole wavelength range, and found the flux (erg s$^{-1}$ cm$^{-2}$) emitted by a quiescent EV Lac in the TESS band to be: F$_{ref} = 2.68 \times 10^{-9} $ erg s$^{-1}$ cm$^{-2}$ . We scaled this with the distance (d) to EV Lac and calculated the energy per flare as follows:

\begin{equation}
    E_{abs} =  \int_{t_0}^{t_1} A(t) dt * F_{ref} * 4\pi d^2,
\end{equation}

where A is the modeled flux of the light curve. The estimated energy $E_{\rm T}$ of each flare in the TESS band is listed in the last column of Table \ref{table:TESS-flares}. We determined 1$\sigma$ uncertainties for each flare from our sampling, described above. The white light FFD of EV Lac observed by TESS is shown in Figure \ref{fig:FFD_TESS}.
\begin{figure} 
   \centering
   \includegraphics[width=0.45\textwidth]{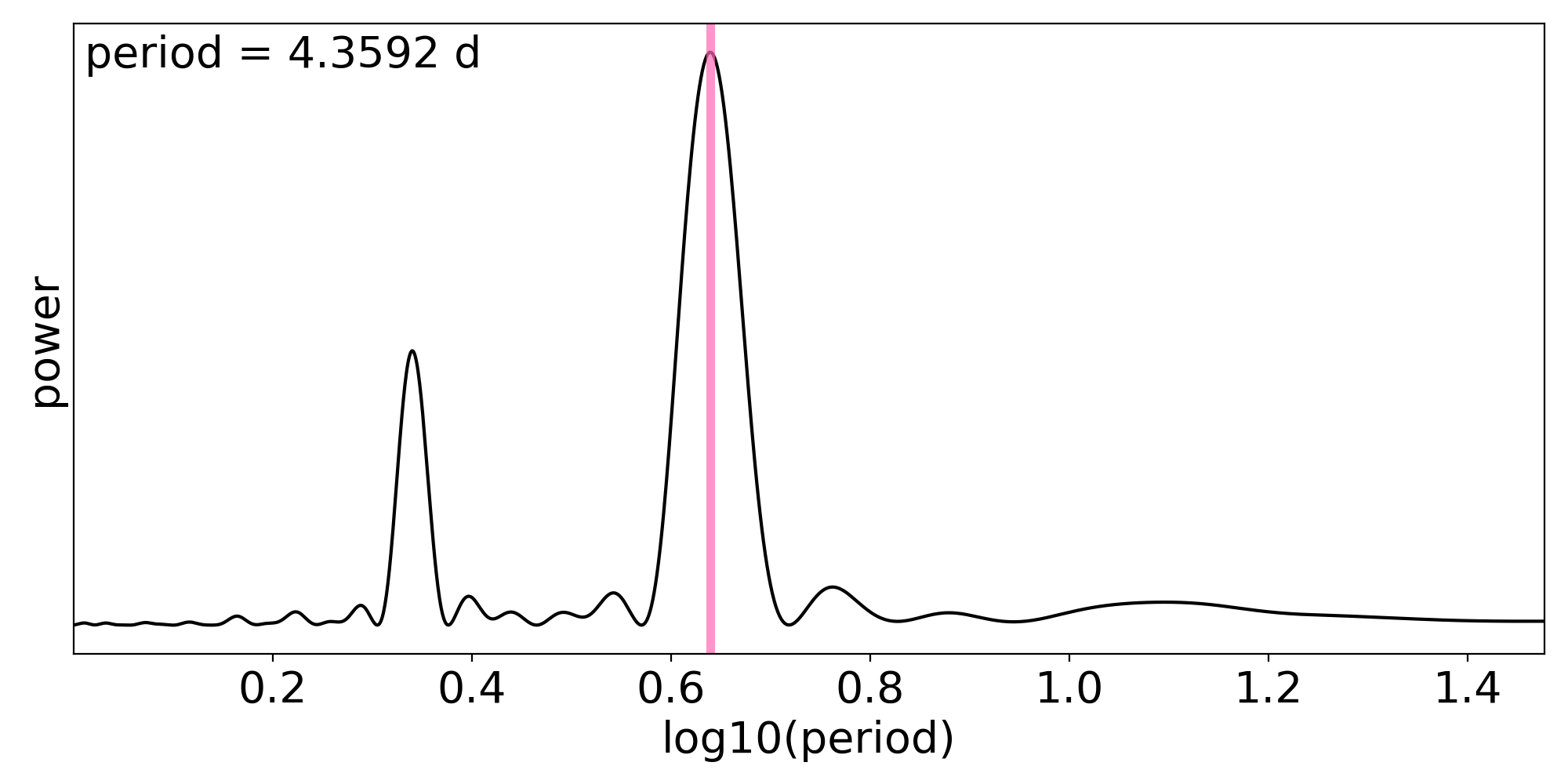}
    \caption{Lomb-Scargle periodogram of the TESS light curve, revealing a prominent peak at the rotation period of EV Lac which we determine to be 4.3592 days.}
\end{figure}
\label{fig:periodogram}

\begin{figure*} 
   \centering
   \includegraphics[width=0.9\textwidth]{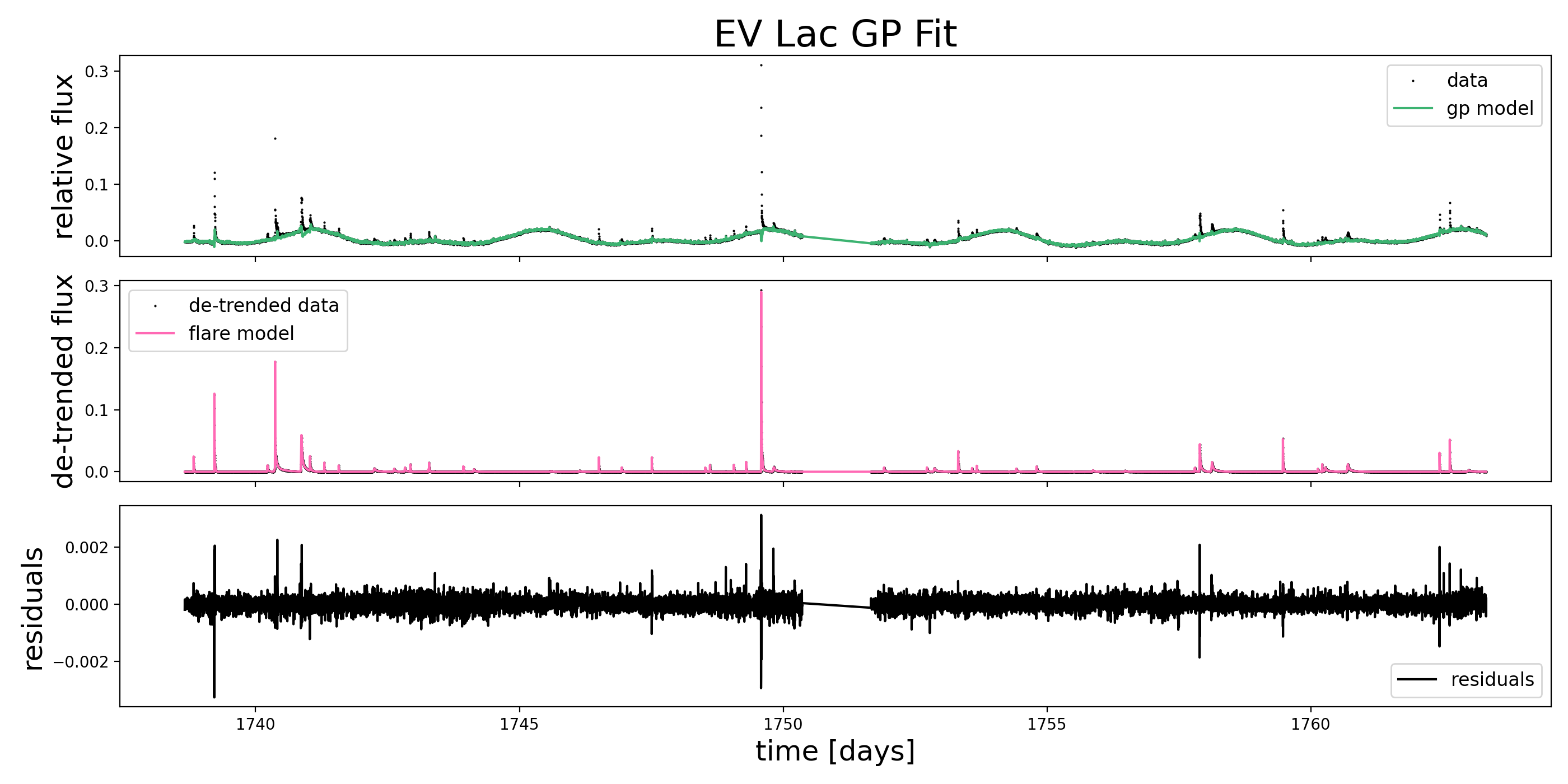}
    \caption{TESS light curve of EV Lac, obtained during Sector 16 (11 Sep, 2019 - 07 Oct, 2019). The top panel shows the light curve (black) and GP fit of the spot modulation and long term variability (green). The middle panel shows the detrended light curve after subtracting the GP model, revealing the flares, with the flare model overplotted in pink. The bottom panel shows the residuals obtained after subtracting both the GP and flare models from the light curve. Our flare analysis identifies 56 EV Lac flares during the 23.2 days covered by the TESS light curve.}
\end{figure*}
\label{fig:tess lc}

\begin{figure} 
  \centering
  \includegraphics[width=0.5\textwidth]{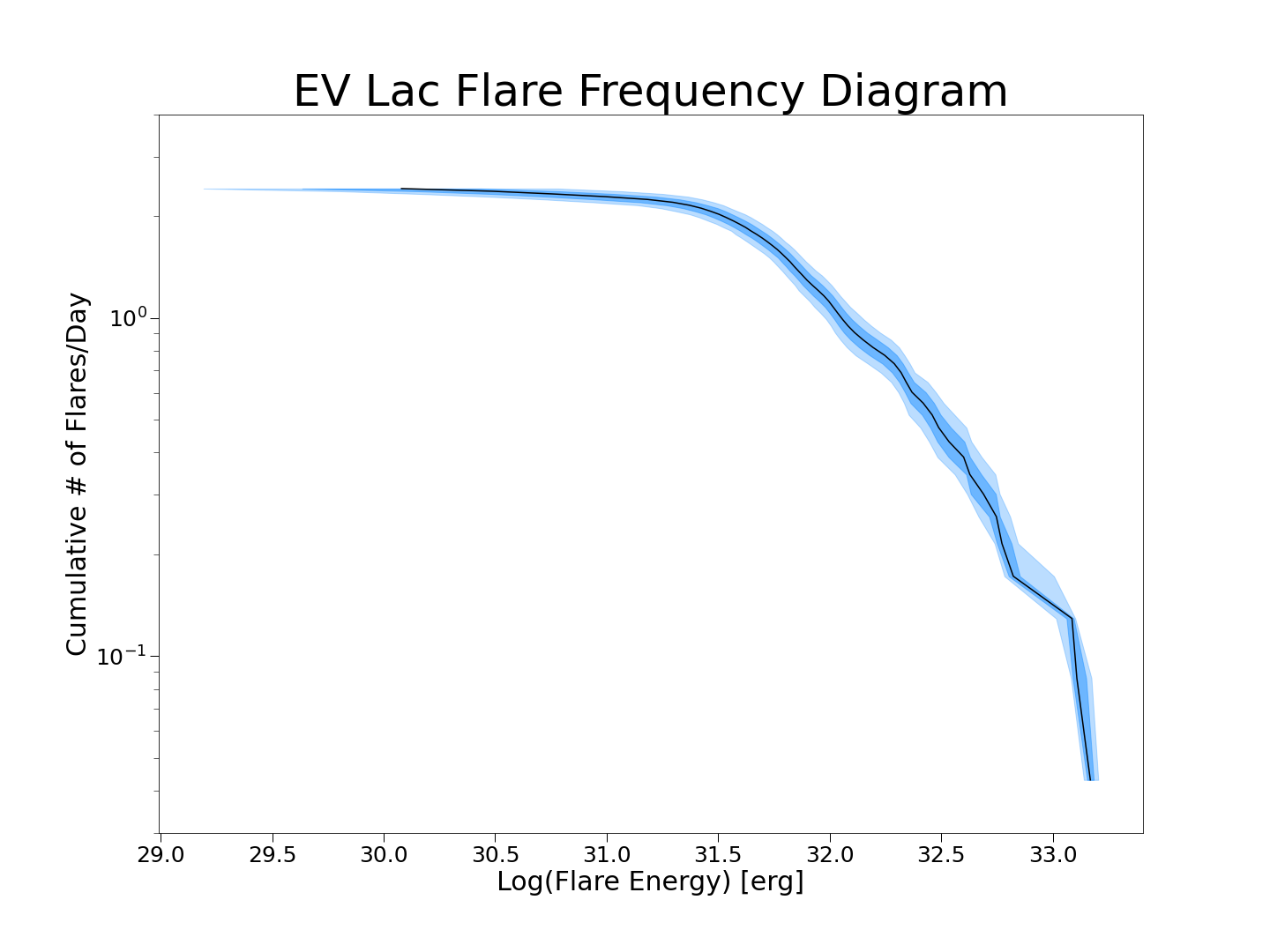}
    \caption{Flare frequency distribution (FFD) of EV Lac and its associated 1$\sigma$ uncertainty (shading) determined from our modeling of the TESS data. The TESS white light flares follow the expected power law distribution from $10^{31.5} - 10^{33}$ ergs. It is harder to detect lower energy flares due to their small amplitudes and durations, resulting in the flattening of the FFD at low energies. Larger flares that occur less frequently may not be seen in the observing baseline of a single TESS sector.}
\end{figure}
\label{fig:FFD_TESS}

\begingroup
\setlength{\tabcolsep}{10pt} 
\renewcommand{\arraystretch}{0.9}
\begin{table*}[!htbp]
    \small
 	\caption{Properties of flares observed by TESS}
 	\label{table:TESS-flares}
     \centering
     \begin{tabular}{cccccc}
     \hline
      flare ID & Peak Time ($T_{0}$) & FWHM & Amplitude & ED & log $E_{\rm T}$ \\
       \hline
        & BJD - 2457000 & min & relative flux & seconds & erg \\
       \hline
        1        &      1738.82433714    &    64      &     0.027674     & 14.1 & 32.06       \\
        2 (T1)      &      1739.21878855    &    202     &     0.122476     & 72.4 & 32.77    \\
        3 (T2)       &      1740.22574895    &    40      &     0.009205     & 16.5 & 32.13      \\
        4        &      1740.36880662    &    162     &     0.174420     & 53.9 & 32.64       \\
        5        &      1740.41186281    &    18      &     0.023720     & 164.6 & 33.13      \\
        6        &      1740.87020271    &    114     &     0.057260     & 188.7 & 33.19     \\
        7        &      1741.02992716    &    148     &     0.024194     & 25.3 & 32.31     \\
        8 (T3)       &      1741.3021531     &    20      &     0.014463     & 6.2 & 31.70      \\
        9        &      1741.57437895    &    102     &     0.011378     & 5.9 & 31.68      \\
        10       &      1742.24938761    &    186     &     0.006970     & 31.7 & 32.41     \\
        11       &      1742.62717002    &    30      &     0.005604     & 11.8 & 31.98     \\
        12       &      1742.82856132    &    74      &     0.007248     & 15.9 & 32.11      \\
        13 (T4)       &      1742.93272922    &    50      &     0.014444   & 5.4 & 31.65 \\
        14       &      1743.27439985    &    4       &     0.004253     & 3.5 & 31.46      \\
        15       &      1743.2882889     &    34      &     0.016011     & 9.2 & 31.88     \\
        16       &      1743.93690724    &    24      &     0.008897     & 4.2 & 31.53     \\
        17       &      1744.1410761     &    44      &     0.004129     & 10.8 & 31.94    \\
        18       &      1745.58970178    &    10      &     0.001775     & 1.0 & 30.92  \\
        19       &      1745.6049797     &    26      &     0.001949     & 2.7 & 31.34  \\
        20       &      1746.13692908    &    24      &     0.002729     & 5.8 & 31.68      \\
        21 (T5)       &      1746.50498793    &    34      &     0.024221     & 9.6  & 31.89    \\
        22       &      1746.93693614    &    96      &     0.007408     & 8.5 & 31.84     \\
        23       &      1747.50916321    &    76      &     0.021192     & 8.1  &  31.82   \\
        24       &      1748.52306041    &    22      &     0.007969     & 4.3  & 31.54    \\
        25       &      1748.61195002    &    68      &     0.011904     & 7.0 & 31.76      \\
        26       &      1749.06473153    &    60      &     0.012007     & 4.3 &  31.55     \\
        27 (T6)       &      1749.29667793    &    122     &     0.013921     & 7.7 & 31.80     \\
        28 (T7)       &      1749.58279152    &    236     &     0.292071     & 158.0 &  33.11     \\
        29       &      1749.59945834    &    4       &     0.017898     & 3.7 & 31.48     \\
        30       &      1749.60501394    &    4       &     0.015053     & 41.0 & 32.52   \\
        31       &      1749.81334917    &    120     &     0.010527     & 39.8 & 32.51       \\
        32       &      1750.14529684    &    6       &     0.002878     & 0.9 & 30.85    \\
        33       &      1751.91198033    &    318     &     0.007173     & 14.7 & 32.08     \\
        34       &      1752.72309347    &    58      &     0.006594     & 13.2 & 32.03     \\
        35       &      1752.8661492     &    120     &     0.006238     & 29.8 & 32.39    \\
        36       &      1753.31614945    &    94      &     0.033520     & 27.0 &  32.34  \\
        37       &      1753.58003832    &    54      &     0.006111     & 6.2 & 31.71    \\
        38       &      1753.66892718    &    6       &     0.009420     & 3.4 & 31.44  \\
        39       &      1753.67726051    &    16      &     0.002388     & 2.1 &  31.23  \\
        40       &      1754.28281555    &    4       &     0.001950     & 0.5 &  30.57  \\
        41       &      1754.41614871    &    104     &     0.006912     & 10.2 & 31.92  \\
        42       &      1754.7994814     &    182     &     0.008560     & 16.3 & 32.12  \\
        43       &      1755.50225756    &    4       &     0.002204     & 0.4 & 30.51        \\
        44       &      1755.87308981    &    40      &     0.003324     & 9.0 & 31.87         \\
        45       &      1756.48142107    &    24      &     0.003474     & 7.0 &  31.76         \\
        46 (T8)       &      1757.80085975    &    38      &     0.006203     & 9.1 & 31.87         \\
        47       &      1757.8966926     &    128     &     0.039899     & 84.2 &  32.84         \\
        48       &      1758.12863584    &    118     &     0.014824     & 72.5 &  32.77        \\
        49 (T9)      &      1759.47168357    &    228     &     0.056124     & 37.3 & 32.48         \\
        50       &      1760.13140146    &    22      &     0.004910     & 6.4 &  31.72         \\
        51       &      1760.21612311    &    56      &     0.009914     & 9.9 &  31.91         \\
        52       &      1760.27584492    &    158     &     0.008755     & 36.7  &  32.48       \\
        53       &      1760.70084196    &    184     &     0.012990     &  59.3 &  32.68          \\
        54       &      1762.44110719    &    108     &     0.032657     & 13.2 & 32.03          \\
        55       &      1762.63416142    &    128     &     0.048114     & 28.9 &  32.37        \\
        56       &      1762.99110355    &    80      &     0.004179     & 20.6  & 32.23        \\

       \hline
      \end{tabular}
\end{table*}
\endgroup

\subsection{\textit{Swift}/{\normalfont XRT}}
The \textit{Swift} XRT light curve is shown in Figure \ref{fig:xrt lc}. We constructed the light curve using \texttt{XSELECT} and present it with a time binning of 120 s. We see a rise in x-rays at $t \sim$1749.3 d. A flare is also seen in the TESS data right after this observation, but there is no direct overlap. We cannot confirm if the rise is due to a flare. In order to estimate the quiescent X-ray flux of the star, we used XSPEC v12.10.1f \citep{1996ASPC..101...17A}, an X-ray Spectral Fitting Package developed by HEASARC \footnote{See https://heasarc.gsfc.nasa.gov/xanadu/xspec/} for spectral fitting.

To prepare spectrum for fitting, we used \texttt{GRPPHA} to bin the XRT spectrum of the source obtained by using \texttt{XSELECT} to have at least 20 counts per bin, a necessary condition to use $\chi^{2}$ statistics \textbf{\citep{2019PASJ...71...75Y}}. We used the two-discrete temperature (2$T$) VAPEC \citep{2001ApJ...556L..91S}) model to fit the observed spectrum. A correction due to column absorption was considered while fitting using $N_{H}$ = 4.0 $\times$ 10$^{18}$ cm$^{-2}$. \textbf{The fitted temperatures are $T_{1}$ = 5.0$^{+1.0}_{-0.6}$ MK and $T_{2}$ = 20.0$^{+7.5}_{-5.2}$ MK, and the corresponding volume emission measures (VEM) are 1.5$^{+0.6}_{-0.6}$ $\times$ 10$^{51}$ cm$^{-3}$ and 0.9$^{+0.3}_{-0.3}$ $\times$ 10$^{51}$ cm$^{-3}$ respectively.} Using this model, we estimate the quiescent X-ray flux of EV Lac to be 1.0 $\times$ 10$^{-11}$ erg cm$^{-2}$ s$^{-1}$ in the 0.3-2.0 keV energy band (soft X-rays).

\textbf{We used 2$T$ \texttt{VAPEC} model to fit the spectrum of event observed in between $t$ = 1749.2 and 1749.3 d during which we observed increase in x-rays. The count rate is 0.76 cnts/s. So we applied pile-up correction before extracting the spectrum. The fitted temperatures are $T_{1}$ = 3.9$^{+1.3}_{-1.0}$ MK and $T_{2}$ = 20.9$^{+12.8}_{-4.1}$ MK, and the corresponding volume emission measures (VEM) are 2.8$^{+1.8}_{-0.9}$ $\times$ 10$^{51}$ cm$^{-3}$ and 3.4$^{+0.9}_{-0.9}$ $\times$ 10$^{51}$ cm$^{-3}$ respectively. We estimate the flux corresponding to this event to be 1.9 $\times$ 10$^{-11}$ erg cm$^{-2}$ s$^{-1}$ in the 0.3-2.0 keV energy band.  }
\begin{figure} 
   \centering
   \includegraphics[width=0.5\textwidth]{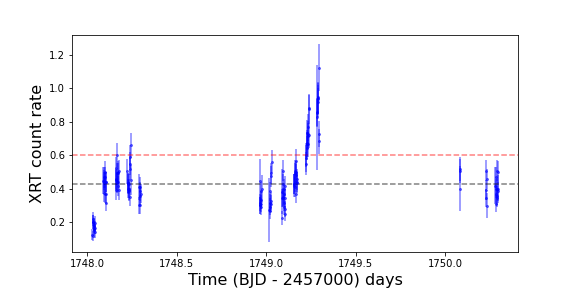}
    \caption{\textit{Swift} XRT light curve of EV Lac with 120 s time bins. The black and red dashed lines correspond to median and median+1$\sigma$ count rates. We cannot confirm that the increase in X-rays at $t\sim$1749.3 d is due to a flare.}
\end{figure}\label{fig:xrt lc}
\subsection{\textit{Swift}/{\normalfont UVOT}}
\begin{figure} 
   \centering
   \includegraphics[width=0.5\textwidth]{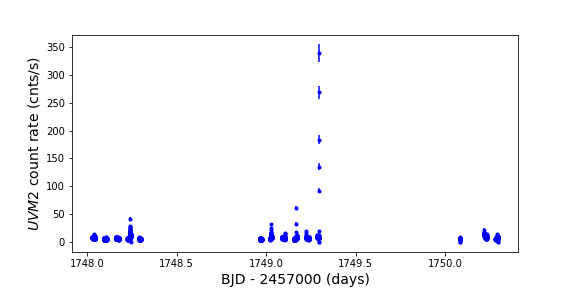}
    \caption{\textit{Swift} UVOT light curve of EV Lac. \textbf{The time binning is 11.033 s.} We identified nine flares in this light curve.}
\end{figure}
\label{fig:uvot light curve}
The \textit{Swift} UVM2 light curve obtained by using the cleaned event list is shown in Figure \ref{fig:uvot light curve} \textbf{with a time binning of 11.033 s}. The median count rate in the light curve is 6.3 counts s$^{-1}$ and corresponds to the quiescent level for the UVM2 filter. We identified nine flares in this light curve. Two flares were observed on 2019 September 21, five on 2019 September 22, and two on 2019 September 23. The full duration of three flares extend beyond the durations of our observations. Hence they were only partially observed. One started at $t$ = 1748.2379 d, the next started at $t$ 1749.2952 d, and another was observed only during its decay phase on $t$ = 1750.21 d. 

We converted the count rate in the UVM2 filter to flux in units of erg cm$^{-2}$ s$^{-1}$ \AA$^{-1}$ by using an average count rate to flux conversion ratio of 8.446 $\times$ 10$^{-16}$. The conversion ratio is part of the {\it Swift} UVOT CALDB and was obtained using GRB models\footnote{\url{https://heasarc.gsfc.nasa.gov/docs/heasarc/caldb/swift/docs/uvot/uvot_caldb_counttofluxratio_10wa.pdf}}. Using this conversion ratio and the FWHM ($\Delta_{uv}$ = 530 \AA) of the UVM2 filter, we found the quiescent UVM2 flux to be 2.8 $\times$ 10$^{-12}$ erg cm$^{-2}$ s$^{-1}$. 

\begin{figure*}
   \centering
   \includegraphics[width=\textwidth]{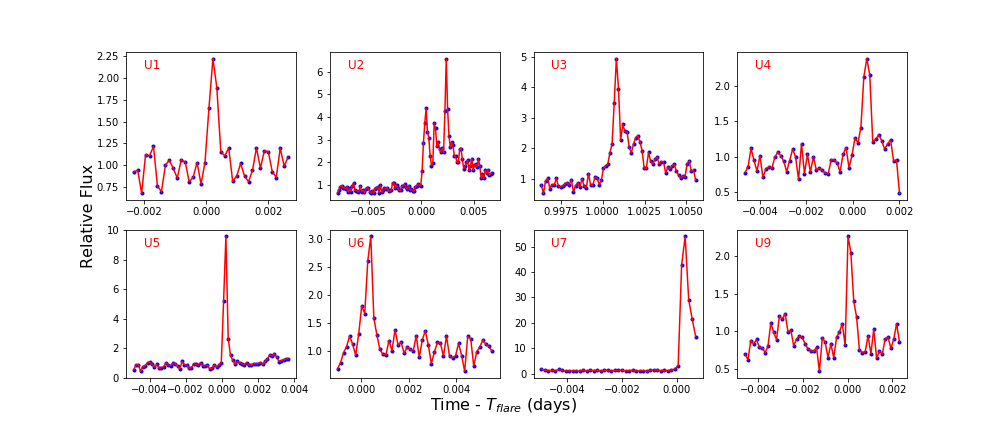}
    \caption{Individual flares observed by \textit{Swift} UVOT. The blue dots in each plot represent the observed fluxes and the red line is the connecting line. The time axis is centered at $T_{0}$ which is the TESS time when a given flare started. The time binning is 11.033 s. One of the UVOT flares, U8, was only observed during the decay phase and is not shown here.}
    \label{fig:flares_uvot}
\end{figure*}

The detailed morphologies of the individual flares identified in the UVOT light curve are shown in Figure \ref{fig:flares_uvot} with a time binning of 11.033 s. The flux plotted along the $Y$-axis of each plot is the relative flux obtained by dividing the flux by the median flux. Likewise, the time plotted along the $X$-axis of each plot is centered at $T_{0}$ which is the TESS time at which a given flare started.  To compute the flare energies, we first estimated the ED of each flare which is the time during which the flare produces same the amount of energy as the star does when it is in its quiescent state \citep{1972Ap&SS..19...75G}. The flare energies $E_{f}$ were then computed by:
\begin{equation}
    E_{f} =  ED \times 4 \pi d_{\star}^{2} \times F_{q}
\end{equation}
where $d_{\star}$ is the distance to the star, and $F_{q}$ is the quiescent flux in units of erg cm$^{-2}$ s$^{-1}$. In Table \ref{table:properties of UVOT flares}, we list the estimated energies of all flares, along with their start times, stop times and EDs. A lower limit on the flare energy is given for each of the three flares which were not observed for their full durations.\

\begin{table*}[htbp]
 	\caption{Properties of flares observed by \textit{Swift} UVOT}
 	\label{table:properties of UVOT flares}
     \centering
     \begin{tabular}{cccccc}
     \hline
      flare ID & time start ($T_{0}$) & time stop & duration & ED & energy \\
       \hline
        & BJD - 2457000 & BJD - 2457000 & min & min & 10$^{30}$ erg \\
       \hline
        U1 & 1748.0401 & 1748.0408 & 0.92 & 0.77 & 0.40 \\
        U2 & 1748.2379 &  & $>$9.6 & $>$16.3 & $>$8.4 \\
        U3 & 1749.0260 & 1749.0314 & 7.7 & 6.7 &  3.4 \\
        U4 & 1749.1059 & 1749.1075 & 2.2 & 1.1 & 0.57 \\
        U5 & 1749.1663 & 1749.1669 & 0.74 & 2.8 & 1.4 \\
        U6 & 1749.2232 & 1749.2238 & 0.92 & 0.93 & 0.48 \\
        U7 & 1749.2952 &  & $>$1.3 & $>$26.0 & $>$13.3 \\
        U8 & & 1750.2256 & $>$11.2 & $>$6.7 & $>$3.4 \\
        U9 & 1750.2913 & 1750.2917 & 0.55 & 0.43 & 0.22 \\
       \hline
      \end{tabular}
\end{table*}

\subsection{NICER}
\label{subsection:NICER analysis}

We show the light curve of EV Lac obtained by NICER in the lower plot of Figure \ref{fig:tess_obs_times}. The median count rate of the light curve is $M_{\rm N}$ = 16.0 counts s$^{-1}$ and the standard deviation is $\sigma_{N}$ = 3.8 counts s$^{-1}$. To obtain these parameters, we first excluded any large events with $>$25 counts s$^{-1}$ after an initial inspection of the light curve, ensuring only the large flare like events were excluded and almost all the quiescent state of the star was included. We then used a threshold cut-off value of $M_{\rm N}$ + 2.5$\sigma_{N}$ to identify flares in the light curve. We identify 14 flares, but none of them were observed for their full duration. In the lower panel of Figure \ref{fig:tess_obs_times}, each flare is given an ID with a letter `N' followed by a number. The black dashed line corresponds to $M_{\rm N}$ and the red dashed line corresponds to $M_{\rm N}$ + 2.5$\sigma_{N}$. The events which showed only a single point brightening were excluded from our flare sample. There is a flare-like event at $t$ = 1759.5 d but it appears to be due to background noise. This is because there is no flux enhancement in the 0.3-2.0 keV energy band during this event, which is not the case during a flare. We note that there is a very weak indication of a feature which might be due to rotational modulation in the NICER light curve in the lower panel of Figure \ref{fig:tess_obs_times}. However, it is not very convincing.    

We used \texttt{XSPEC v12.10.1f} for spectral fitting of the flares observed by NICER. Before fitting, we binned the spectra by using \texttt{GRPPHA}. We used the same Response Matrix File (RMF) and Ancillary Response File (ARF) described in Section \ref{subsection:NICER data} during spectral fitting of NICER flares. We used a three temperature (3$T$) \texttt{VAPEC} model to fit the spectrum of each flare except for flare N6, together with the abundances of \cite{1989GeCoA..53..197A}. We used the F-test to compare two temperature (2$T$) and three temperature \texttt{VAPEC} models, and found that 3$T$ \texttt{VAPEC} model gives a better fit except for flare N6. We applied a correction due to interstellar absorption to each flare by using a fixed value of column density $N_{H}$ equal to 4 $\times$ 10$^{18}$ cm$^{-2}$.  A similar value was used by \cite{2005ApJ...621..398O} for this star. The values of fitted parameters are listed in Table \ref{table:NICER XRT}. \textbf{We report the errors of fitted parameters at 90\% confidence level.} A discussion of coronal abundances is deferred to Section \ref{subsec:FIP_IFIP}. In Table \ref{table:NICER XRT}, $T_{i}$ (i=1,2,3) are the fitted flare temperatures, and EM$_{i}$ (i=1,2,3) are the corresponding volume emission measures. \textbf{Likewise, `F-test prob.' is the probability of F-test which is used to compare 2$T$ and 3$T$ \texttt{VAPEC} models. A lower probability implies a significant  improvement in the fit due to addition of a component in the 2$T$ \texttt{VAPEC} model.} The last column `Quies.' corresponds to the fitted values for the quiescent level of the star. The values of flare fluxes are listed in Table \ref{table:flare energies NICER}. 

In Figure \ref{fig:NICER flareN2}, we show an example of a flare spectrum fitted by using 3$T$ \texttt{VAPEC} model. The upper panel shows the data and model (solid line), and the lower panel shows the residuals. The spectra of all other flares observed by NICER and the quiescent level are available online.  

\begin{figure} 
   \centering
   \includegraphics[width=0.5\textwidth]{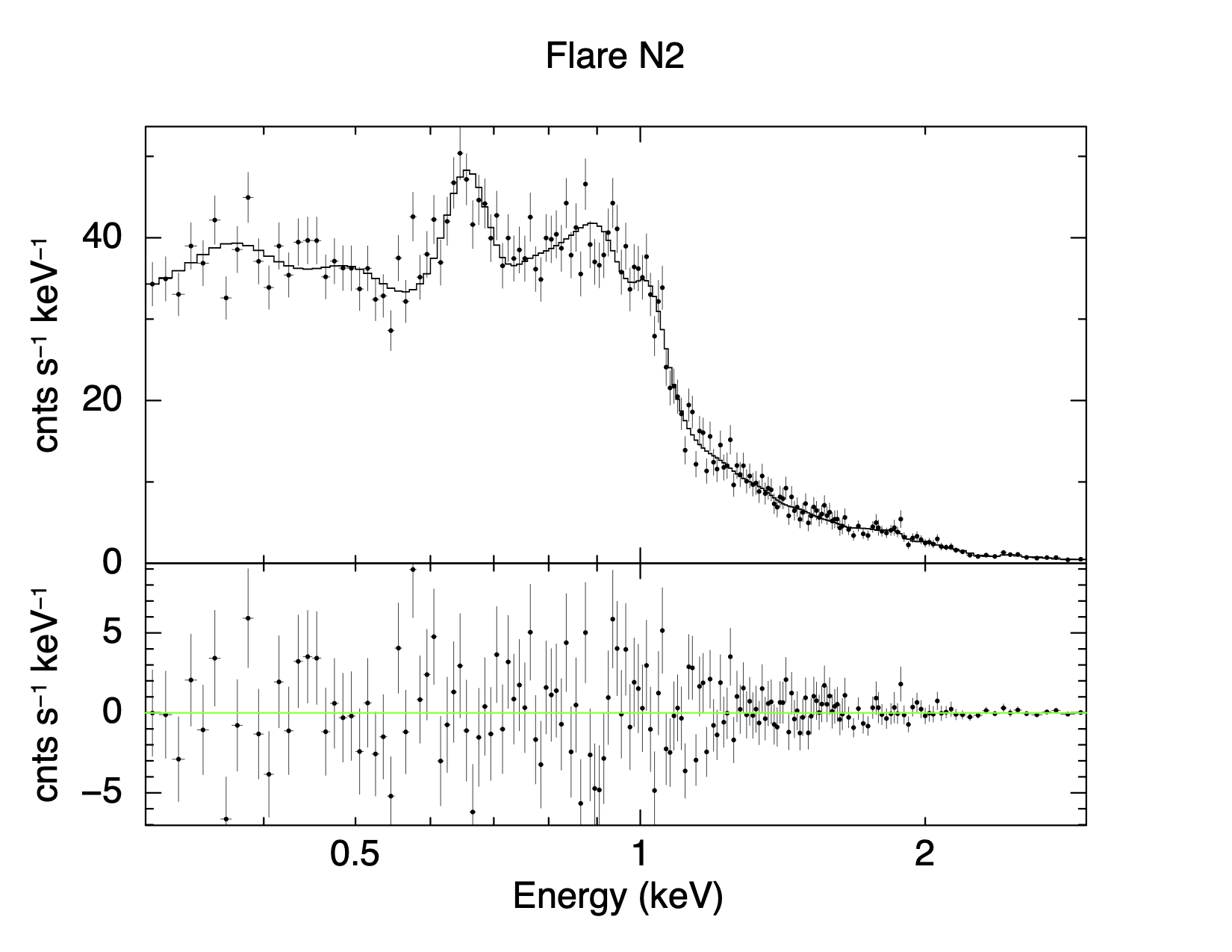}
    \caption{\textbf{An example of fitting of an X-ray flare spectrum by using \texttt{xspec}. The upper panel shows the data and 3$T$ VAPEC model (solid line), and the lower panel shows the residuals.}}
\end{figure}
\label{fig:NICER flareN2}

We estimated the X-ray energy of each NICER flare by multiplying the flux obtained by fitting each flare by 4$\pi d_{\star}^{2}$ and by the duration of each flare observed by NICER. We list the energies of each flare in Table \ref{table:flare energies NICER}. In this table, the first column is the flare ID, second column is the flare start time. The third column gives information about the rise or decay phase of a given flare when it was observed. In the case of five flares: N1, N2, N6, N9 and N13, we only see enhancement in X-ray flux with almost a constant value during the observation.

The fourth column is the X-ray band in which the flare had \textbf{a} significant number of counts. In the fifth, sixth and seventh columns, we list three different values of duration for each flare: i) $t_{N}$ is the total time for which a given flare was observed by NICER, ii) $t_{T}$ is the duration of a given flare in TESS data if it was observed simultaneously by TESS, and iii) $T_{N,gaps}$ is the total possible duration of a given flare by considering the total time in between two consecutive quiescent levels that were observed by NICER, with flare enhancement in between them. 

The energies corresponding to each time duration are given in the last three columns. Since the flares were not observed for their full duration, we can only estimate the maximum and minimum values of energies in terms of duration. Energies corresponding to $t_{N}$ are the minimum values as they correspond to the energies released during the NICER exposure times of the flares. The energies corresponding to $T_{N,gaps}$ are the maximum values for a given flare in terms of the flare duration only. The flare energies corresponding to $t_{N}$ and $T_{N,gaps}$ are estimated by using the same average flare flux we observed during the flares. We did not use any flare model to estimate these energies. These energies are just the rough estimates and do not represent the total energies released during the corresponding flares. They are estimated to see how large the energies might be if we consider the two times $t_{N}$ and $T_{N,gaps}$. 

In Table \ref{table:flare energies NICER}, $F_{f}$ is the average flare flux estimated after subtracting the quiescent flux and $F_{f}$/$F_{q}$ is the ratio of flare flux to quiescent flux for a given flare. Using the NICER light curve, we estimate the quiescent flux of EV Lac to be equal to 1.3 $\times$ 10$^{-11}$ erg cm$^{-2}$ s$^{-1}$.  
\subsubsection{Possibility of a large, complex X-ray flare between $t$ = 1742.10 and 1742.60 d}
NICER observed three enhancements in X-ray flux from EV Lac, in between $t$ = 1742.10 and 1742.30 d, which can be seen in Figure \ref{fig:nicer large flare}. We notice a flare decay phase during the enhancement at $t$ = 1742.20 d, so it is likely that the two enhancements at $t$ = 1742.15 and $t$ = 1742.20 are the decay phases of the same flare. It is also possible that the three enhancements observed at $t$ = 1742.15, 1742.20 and 1742.28 d are parts of a complex flare with two peaks. The total exposure time during the three enhancements is 2820 s, and the total time including the gaps between the flares is 12700 s (3.5 hr). Furthermore, the upper limit in total duration of this complex flare is 12.1 hr. The X-ray energy emitted during the exposure time is log $E$ (erg) = 32.6. No optical flares were observed by TESS during these times.
\begin{table*}
 	\caption{NICER flares spectral fit results}
 	\centering
 	\tabcolsep=0.11cm
     \begin{tabular}{c|ccccccccc}
\hline
flare ID & N0 & N1  & N2 & N3 & N4(peak) & N5 & N6 & N7  & N8   \\ 
\hline
X-ray band (keV)  & 0.3-3.0 & 0.3-2.0 & 0.3-3.0 & 0.3-3.0  & 0.3-3.0  & 0.3-3.0 & 0.3-2.0  & 0.3-3.0 & 0.3-3.0 \\ 
$T_{1}$ (MK) & 3.0$^{+0.5}_{-0.4}$  & 2.8$^{+0.5}_{-0.4}$ & 3.5$^{+2.0}_{-1.5}$   & 3.1$_{-0.1}^{+0.2}$  & 4.8$^{+1.3}_{-1.7}$  & 5.7$_{-1.0}^{+0.9}$  & 5.2$_{-0.7}^{+1.4}$  & 2.9$_{-0.2}^{+0.2}$  & 3.5$_{-0.7}^{+1.8}$   \\ 
EM$_{1}$ (10$^{51}$ cm$^{-3}$) & 1.2$^{+0.3}_{-0.3}$  & 1.2$^{+0.3}_{-0.3}$ & 1.2$^{+0.6}_{-0.6}$   & 1.2$_{-0.3}^{+0.3}$  & 1.8$_{-1.2}^{+1.2}$  & 3.7$_{-0.9}^{+0.9}$  & 3.1$_{-1.2}^{+1.5}$  & 1.2$_{-0.3}^{+0.3}$  & 1.1$_{-0.6}^{+0.3}$     \\  
$T_{2}$ (MK) & 9.2$^{+0.4}_{-0.5}$  & 7.5$^{+1.9}_{-1.4}$ & 9.3$^{+1.6}_{-0.6}$   & 7.9$_{-0.5}^{+0.6}$  & 9.4$_{-1.3}^{+2.1}$  & 11.6$_{-0.8}^{+0.7}$ & 11.4$^{+0.8}_{-0.9}$ & 8.7$_{-0.7}^{+0.5}$  & 8.5$_{-0.8}^{+0.6}$    \\ 
EM$_{2}$ (10$^{51}$ cm$^{-3}$) & 3.7$^{+0.9}_{-0.9}$  & 1.1$^{+0.6}_{-0.3}$ & 6.1$_{-1.8}^{+2.7}$   & 2.9$_{-0.6}^{+0.9}$  & 3.7$_{-0.9}^{+1.2}$  & 7.6$_{-1.8}^{+1.8}$  & 6.4$_{-1.5}^{+1.8}$  & 2.4$_{-0.6}^{+0.9}$  & 3.4$_{-0.9}^{+1.2}$   \\ 
$T_{3}$ (MK)  & 24.4$^{+2.6}_{-1.5}$ & 24.4$^{+12.8}_{-5.8}$ & 25.5$^{+13.1}_{-3.6}$ & 22.0$_{-2.2}^{+2.4}$ & 23.2$_{-1.3}^{+1.3}$ & 29.0$_{-2.9}^{+4.2}$ & --   & 24.4$_{-2.1}^{+3.0}$ & 22.0$_{-2.2}^{+2.3}$   \\ 
EM$_{3}$ (10$^{51}$ cm$^{-3}$) & 11.0$^{+0.6}_{-0.6}$ & 2.3$^{+0.3}_{-0.6}$  & 4.3$^{+0.9}_{-1.2}$   & 2.8$_{-0.3}^{+0.3}$  & 9.0$_{-1.8}^{+0.6}$  & 15.6$_{-1.8}^{+1.5}$ &  - & 4.3$_{-0.3}^{+0.3}$  & 4.3$_{-0.6}^{+0.3}$  \\  
reduced $\chi^{2}$, dof & 1.1, 206 & 1.2, 97 & 0.97, 164  & 1.1, 176 & 1.1, 214 & 1.0, 242 & 0.95, 113  & 1.0, 169  & 1.0, 180  \\
F-test prob. & 2.6e-06 & 2.1e-06 & 0.02 & 4.2e-08 & 2.2e-04 & 4.6e-08 & 0.95 & 2.9e-09 & 4.9e-4 \\
\hline
flare ID & N9 & N10  & N11 & N12 & N13 & Quies.  \\ 
\hline
X-ray band (keV) & 0.3-3.0   & 0.3-3.0  & 0.3-3.0 & 0.3-3.0  & 0.3-3.0 & 0.3-2.0 \\
$T_{1}$ (MK) & 3.0$_{-0.2}^{+0.4}$ & 4.9$_{-0.5}^{+0.7}$  & 3.0$_{-0.4}^{+0.4}$    & 3.0$_{-0.2}^{+0.1}$  & 3.1$_{-0.4}^{+0.7}$  & 2.9$_{-0.1}^{+0.1}$    \\
EM$_{1}$ (10$^{51}$ cm$^{-3}$) & 1.6$_{-0.3}^{+0.3}$   & 3.4$_{-0.6}^{+0.6}$  & 0.8$_{-0.3}^{+0.3}$    & 1.1$_{-0.3}^{+0.3}$  & 1.5$_{-0.6}^{+0.6}$  & 1.0$_{-0.1}^{+0.1}$ \\
$T_{2}$ (MK) &  7.7$_{-0.6}^{+0.8}$   & 11.5$_{-0.8}^{+0.7}$  & 8.8$_{-0.9}^{+0.7}$    & 9.2$_{-0.5}^{+0.4}$  & 7.5$_{-1.6}^{+1.3}$ & 7.8$_{-0.1}^{+0.1}$ \\
EM$_{2}$ (10$^{51}$ cm$^{-3}$) & 2.9$_{-0.6}^{+0.9}$   & 7.3$_{-1.8}^{+2.1}$  & 1.7$_{-0.6}^{+0.9}$    & 2.8$_{-0.6}^{+0.6}$  & 1.5$_{-0.6}^{+0.9}$   & 1.9$_{-0.2}^{+0.2}$   \\
$T_{3}$ (MK) & 24.4$_{-1.7}^{+2.0}$  & 23.2$_{-2.4}^{+4.8}$ & 29.0$_{-4.2}^{+5.3}$ & 29.0$_{-2.9}^{+3.2}$ & 27.8$_{-3.8}^{+4.6}$ &  20.9$_{-1.6}^{+2.3}$  \\
EM$_{3}$ (10$^{51}$ cm$^{-3}$) & 5.8$_{-0.3}^{+0.3}$   & 8.9$_{-1.8}^{+1.5}$ & 3.7$_{-0.3}^{+0.3}$ & 4.8$_{-0.3}^{+0.3}$  & 7.3$_{-0.6}^{+0.6}$ & 0.70$_{-0.1}^{+0.1}$   \\
reduced $\chi^{2}$, dof & 1.1, 190 & 1.1, 213   & 0.83, 157 & 1.0,208 & 1.0, 137   & 1.2, 157  \\
F-test prob. & 3.1e-27 & 1.5e-08 & 1.2e-06 & 6.4e-12 & 4.0e-4 & 7.6e-16 \\
\hline
\end{tabular}
\end{table*}
\label{table:NICER XRT}
\begin{table*}
\scriptsize
 	\caption{Properties of flares observed by NICER}
     \centering
     \begin{tabular}{c|cccccccccccc}
     \hline
       Flare ID & flare time & phase & X-ray band & \multicolumn{3}{c}{duration} & av. flux & $F_{f}$/$F_{q}$ & \multicolumn{3}{c}{$Energy$}\\
       \hline
       & & & & $t_{\rm N}$ & $t_{\rm T}$ & $t_{\rm N,gaps}$ & 10$^{-11}$ & & log $E_{\rm min,N}$ & log $E_{\rm T}$ & log $E_{\rm N,gaps}$ \\
       \hline
        & BJD - 2457000 & & keV & sec & sec & hr & erg cm$^{-2}$ s$^{-1}$ & &  erg & erg & erg \\
       \hline
       N0 & 1739.25 & decay & 0.3-3.0 & 574 & 5400 & 10.6 & 5.1 & 3.9 & 32.0 & 33.0 & 33.8 \\
       N1 & 1740.22 & rise & 0.3-2.0 & 219 & 1814 & 1.5 & 0.5 & 0.4 & 30.5 & 31.4 & 31.9 \\
        N2 & 1741.11 & -- & 0.3-3.0 & 470 & & 21.3 & 2.6 & 2.0 & 31.6 & & 33.8 \\
       N3 & 1741.31 & decay & 0.3-3.0 & 952 & 950 & 1.5 & 1.5 & 1.2 & 31.6 & 31.6 & 32.4 \\
       N4 & 1742.15 & peak/decay & 0.3-3.0 & 1864 & & 3.1 & 2.5 & 1.9 & 32.1 & & 32.9 \\
        N5 & 1742.28 & decay & 0.3-3.0 & 956 & & 9.0 & 7.5 & 5.8 & 32.3 & & 33.9 \\
        N6 & 1742.86 & -- & 0.3-2.0 & 239 & & 4.4 & 1.7 & 1.3 & 31.1 & & 32.9 \\
        N7 & 1742.92 & rise & 0.3-3.0 & 910 & 1080 & 1.6 & 1.9 & 1.5 & 31.7 & 31.8 & 32.5 \\
        N8 & 1746.6 & peak/decay & 0.3-3.0 & 731 & 1443 & 1.6 & 2.1 & 1.6 & 31.7 & 32.4 & 32.6 \\
       N9 & 1749.31 & -- & 0.3-3.0 & 756 & 1555 & 7.5 & 2.4 & 1.9 & 31.7 & 32.1 & 33.3 \\
       N10 & 1749.63 & decay & 0.3-3.0 & 839 & 11163 & 7.5 & 5.5 & 4.2 & 32.2 & 33.3 & 33.7 \\
       N11 & 1750.21 & rise & 0.3-3.0 & 489 & & 3.0 & 1.3 & 1.0 & 31.3 & & 32.6 \\
       N12 & 1757.82 & peak/decay & 0.3-3.0 & 1088 & 1322 & 19.8 & 2.3 & 1.8 & 31.9 & 32.0 & 33.7 \\
       N13 & 1759.51 & -- & 0.3-3.0 & 230 & 6005 & 1.4 & 2.7 & 2.1 & 31.3 & 32.7 & 32.6 \\
       \hline
      \end{tabular}
\end{table*}
\label{table:flare energies NICER}
\begin{figure} 
   \centering
   \includegraphics[width=0.5\textwidth,height=0.3\textwidth]{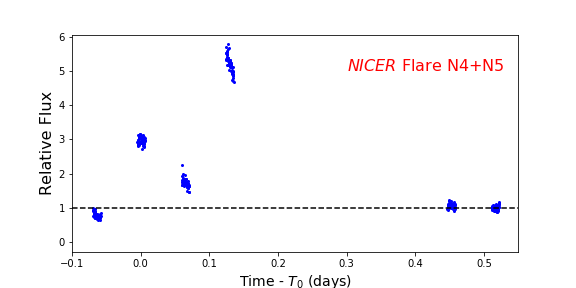}
    \caption{Flares observed by NICER in between $t$ = 1742.10 d and $t$ = 1742.60 d. The black dashed line corresponds to the quiescent level ($M_{N}$). The time along the $X$-axis is centered at $T_{0}$ = 1742.15 d. }
\end{figure}
\label{fig:nicer large flare}
%
%
%
\subsection{{\normalfont LCOGT} light curve}
The EV Lac light curve obtained by LCOGT is shown in Figure \ref{fig:LCOGT lc}. The subplot in the left shows the light curve which was obtained on 2019 September 17 and that in the right shows the one obtained on 2019 September 21. As seen in the figure, one full flare was observed by LCOGT at $t$ = 1749.89 d, and the observed peak flux was $\sim$20\% brighter than quiescence. Only NICER observed the star simultaneously during the time of this flare. However, we do not notice a clear flare-like event in the NICER light curve during that time. There is a very slight enhancement in X-ray level with respect to the median value ($M_{N}$) but is within the $M_{N}$+2.5$\sigma$$_{N}$ value. So it is hard to decide if it is due to flare or other factors such as instrumental effects. Fluctuations in the quiescent level can be seen in the light curve during other times as well. The LCOGT flare occurred during the TESS data downlink time. As a result, we have no information from TESS about this event. We might have observed the decay phase of a flare in the light curve shown in the left sub-plot. However, we do not see a flare during the same time in TESS light curve.

We estimate the U band flare energy in an analogous manner to the UVOT flares. We compute the equivalent duration (827 s) of the flare by first normalizing the light curve with a linear fit (masking the time span of the flare), integrating over the 13-minute period (792 s) that includes the flare rise time and the decay (until it becomes indistinguishable from the quiescent flux level). A U band magnitude measurement is unavailable for EV Lac in the literature, so we estimate it based off the B band magnitude (11.85 mag; \citealt{2012yCat.1322....0Z}) and the U-B = 1.22 color for M4 dwarfs from \cite{2013ApJS..208....9P}. Using the published zero-point magnitude flux for the U band and mean width of the filter \citep{1998A&A...333..231B} and distance to EV Lac, we estimate EV Lac's quiescent luminosity in the U band to be 4.6$\times$10$^{28}$ erg s$^{-1}$. With the equivalent duration measured from the light curve, we find the U band energy of the flare to be 3.8$\times$10$^{31}$ erg. 

\begin{figure} 
   \centering
   \includegraphics[width=0.5\textwidth,height=0.3\textwidth]{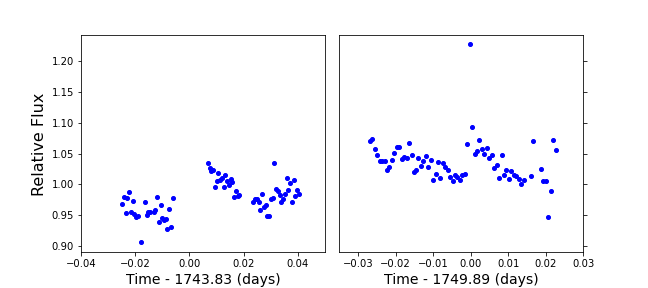}
    \caption{Light curve of EV Lac obtained by LCOGT, on 17 Sep, 2019 (left) and on 21 Sep, 2019. A large flare was observed at $t$ = 1749.89 d.}
\end{figure}
\label{fig:LCOGT lc}
\subsection{Flare observed simultaneously by {\normalfont TESS}, \textit{Swift}/{\normalfont UVOT} and {\normalfont NICER}}
\begin{figure}
    \centering
   \includegraphics[width=0.45\textwidth,height=0.4\textwidth]{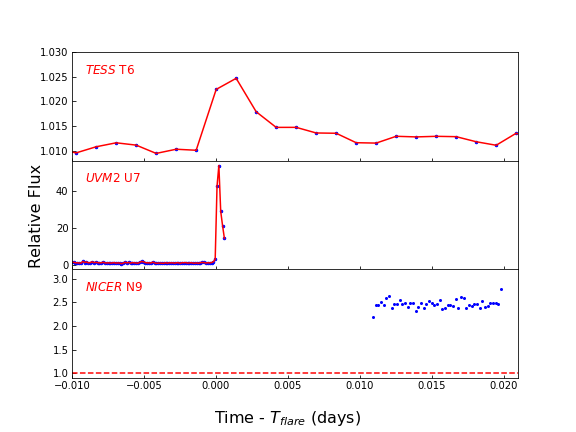}
    \caption{Flare observed simultaneously by TESS, \textit{Swift}/UVOT and NICER. Note that the cadence length of TESS data is 2.0 min, \textit{Swift}/UVOT is 11.033 s and that of NICER data is 5.0 s. The red dashed line in the lowermost subplot corresponds to the quiescent level ($M_{N}$) in the NICER light curve.}
    \label{fig:flare_TESS_UVOT_nicer}
\end{figure}

The flare T6 observed by TESS at $t$ = 1749.29 d was partly observed simultaneously by \textit{Swift}/UVOT (ID: U7) as well as by NICER (ID: N9). The total duration of this flare in the TESS band is 25.9 min. \textit{Swift}/UVOT observed only the rise phase and the initial decay phase of the flare for 1.3 min. NICER observed only a part of the decay phase during which X-ray flux was almost constant for 12.6 min. This flare is shown in Figure \ref{fig:flare_TESS_UVOT_nicer}. In the TESS band, the energy of this flare is estimated to be equal to log $E_{\rm T}$ (erg) = 31.8, and in UVM2 band, the energy is estimated to be log $E_{\rm U}$ (erg) $>$31.1. Likewise, the estimated energy of this flare in NICER band is log $E_{\rm N}$ (erg) $>$ 31.7.
\subsection{Flare observed simultaneously by \textit{Swift}/{\normalfont UVOT} and {\normalfont NICER}}
\begin{figure}
   \centering
   \includegraphics[width=0.45\textwidth,height=0.4\textwidth]{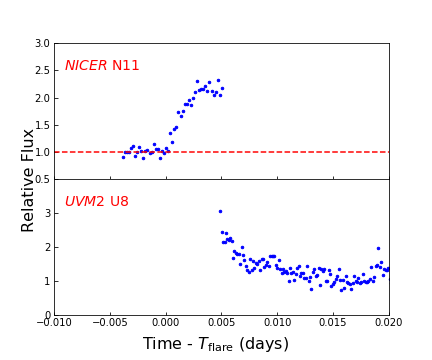}
    \caption{Flare observed simultaneously by NICER and \textit{Swift} $UVOT$. The cadence length of UVOT data is 11.033 s and that of NICER is 5.0 s.  The red dashed line in the subplot showing NICER flares correspond to the quiescent level ($M_{\rm N}$) in NICER light curve.}
    \label{fig:flare_nicer_uvot_simul}
\end{figure}

In addition to U7/N9, another flare U8/N11, shown in Figure \ref{fig:flare_nicer_uvot_simul}, was observed simultaneously by \textit{Swift}/UVOT and NICER. However, neither mission observed the flare for its full duration. NICER observed it during the rise phase for 8.2 min, and UVOT observed it during the decay phase for 11.2 min. The estimated energy of this flare is log $E_{\rm N}$ (erg) $>$ 31.3 in the NICER band and log $E_{\rm U}$ (erg) $>$30.5 in the UVM2 band. This flare occurred during the TESS data downlink period, so we do not have its TESS light curve.

The flares U1, U2, U3, U4, U5, U6 and U9 were observed during data gaps in the NICER light curve. Likewise, no flares were observed simultaneously by TESS and $Swift$ UVOT except the flare T6. There is a weak indication of flux enhancement in \textbf{the} TESS light curve during U2 but it is within the noise level of the light curve, and hence is not detected by \texttt{bayesflare}.
\subsection{Flares observed simultaneously by {\normalfont TESS} and {\normalfont NICER}}
\begin{figure*}
   \centering
   \includegraphics[width=\textwidth,height=0.4\textwidth]{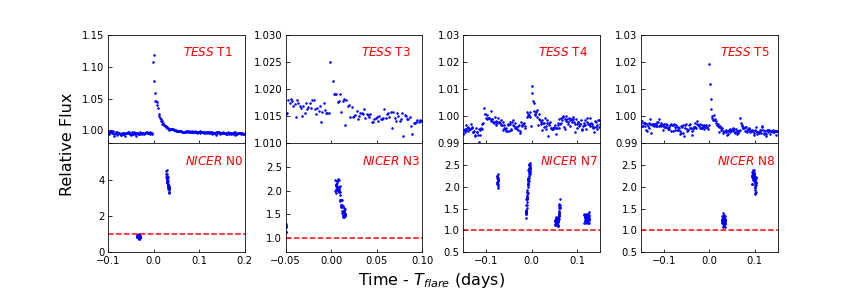}
   \includegraphics[width=\textwidth,height=0.4\textwidth]{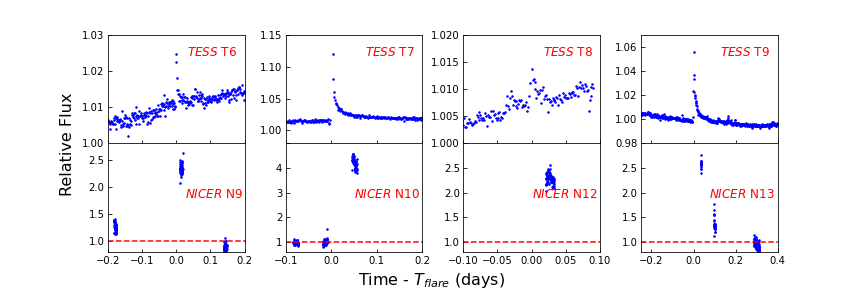}
    \caption{Flares observed simultaneously by TESS and NICER. The red dashed line in the subplots showing NICER flares correspond to the quiescent level ($M_{\rm N}$) in NICER light curve. The cadence length of TESS data is 2.0 min and that of NICER is 5.0 s.}
    \label{fig:flare_TESS_nicer1}
\end{figure*}

Nine flares were observed simultaneously by TESS and NICER. Eight of them are shown in Figure \ref{fig:flare_TESS_nicer1}. The remaining flare is N1 which was observed during its initial rise phase simultaneously for 219 s with NICER. The flare ID and the mission name are mentioned inside each subplot. The red dashed line in the subplots showing NICER flares corresponds to the quiescent level ($M_{\rm N}$) in the NICER light curve. 

In Table \ref{table:tess_nicer energy comparison}, we compare the TESS flare energies $E_{T}$ as well as corresponding bolometric flare energies $E_{\rm bol}$ with the NICER flare energies for the above nine flares. Because of the incomplete information about the flares in NICER data, we can only estimate an upper limit on the ratios $E_{T}$/$E_{N}$ and $E_{\rm bol}$/$E_{N}$. An estimation of X-ray flare energies by considering the duration of corresponding flares observed in TESS data is listed in Table \ref{table:flare energies NICER} in column `log $E_{\rm T}$'.
\addtolength{\tabcolsep}{-3pt}
\begin{table}
 	\caption{Comparison of flare energies $E_{\rm T}$ and $E_{\rm N}$ }
 	\label{table:tess_nicer energy comparison}
     \centering
     \begin{tabular}{ccccc}
     \hline
       Flare ID & log $E_\textrm{T}$ & log $E_\textrm{bol}$ & log $E_\textrm{N}$ & $(E_\textrm{T}/E_\textrm{N})_\textrm{max}$   \\
       \hline
        & erg & erg & erg & \\
        \hline
       T1/N0 & 32.8 & 33.6 & 32.0 & 6.3  \\
       T2/N1 & 32.1 & 32.9 & 30.5 & 40.0  \\
       T3/N3 & 31.7 & 32.5 & 31.6 & 1.3  \\
       T4/N7 & 31.7 & 32.5 &  31.7 & 1.0  \\
       T5/N8 & 31.9 & 32.7 & 31.7 & 1.6  \\
       T6/N9 &31.8 & 32.6 & 31.7 & 1.3  \\
       T7/N10 &33.1 & 33.9 & 32.2 & 8.0  \\
       T8/N12 & 31.9 & 32.7 & 31.9 & 1.0  \\
       T9/N13 & 32.5 & 33.3 & 31.3 & 15.8  \\
       \hline
      \end{tabular}
\end{table}
\addtolength{\tabcolsep}{3pt}

\subsection{Comparison of quiescent luminosity in various bands to bolometric luminosity}
In Table \ref{table:flux comparison}, we list the values of quiescent fluxes of EV Lac which we estimated for various bands in which the star was observed. In addition, we compare those fluxes with the flux in the TESS band. Such ratios will be helpful to estimate the total energy output of a TESS target with comparable spectral type and age as EV Lac.
\begin{table}
 	\caption{Quiescent luminosity in various bands}
 	\label{table:flux comparison}
     \centering
     \begin{tabular}{ccc}
     \hline
       Band & log $L_{\rm Q}$ & log $L_{\rm Q}$/$L_{\rm bol}$ \\
       \hline
        & erg s$^{-1}$ &  \\
       \hline
       TESS & 30.9 & -0.79 \\
       \textit{Swift} XRT & 28.5 (0.3-2.0 keV) & -3.2 \\
       \textit{Swift} UVM2 & 27.9 & -3.8 \\
       NICER & 28.6 (0.3-2.0 keV) & -3.1 \\
       LCO & 28.7 & -3.0 \\
     \hline
      \end{tabular}
\end{table}
\subsection{Analysis of the {\normalfont FIP/IFIP} effect}
\label{subsec:FIP_IFIP}
In the case of active regions in the solar corona, the elements with low-First Ionization Potential (FIP) are found to be more abundant than those with high-FIP (above $\sim$10 eV) when compared to their photospheric abundances. This is known as the FIP effect \citep{1992PhyS...46..202F}. While certain stars also show evidence of a solar-like FIP effect, some active stars also show an Inverse FIP (IFIP) effect where the high-FIP elements are more abundant compared to the low-FIP elements in the corona relative to their photospheric abundances \citep{2001A&A...365L.324B}. Studies of flares on different stars show mixed results regarding how the flares affect the FIP pattern. \cite{1999ApJ...511..405G,2001A&A...365L.318A,2003A&A...411..509R} reported an increase in low-FIP abundances during flares on UX Ari, HR 1099, and AT Mic respectively. However, \cite{2003ApJ...582.1073O}, \cite{2004A&A...416..713G} and \cite{2007MNRAS.379.1075R} did not observe such an effect during flares from other targets: $\sigma^{2}$ Coronae Boraelis, Proxima Centauri and YZ CMi.\\
\begin{table*}
\scriptsize
 	\caption{Abundances measured relative to Fe during a flare compared to the quiescent state}
     \centering
     \begin{tabular}{c|ccccccccc}
     \hline
       El & Quies. & \multicolumn{2}{c}{Flare N0} & 
       \multicolumn{2}{c}{Flare N2} & \multicolumn{2}{c}{Flare N4} & \multicolumn{2}{c}{Flare N5} \\
       \hline
       & X/Fe & X/Fe & Flare/Quies. & X/Fe & Flare/Quies. & X/Fe & Flare/Quies. & X/Fe & Flare/Quies.  \\
       \hline
       O & 0.56$_{-0.02}^{+0.02}$ & 0.73$_{-0.09}^{+0.11}$& 1.30$_{-0.17}^{+0.20}$ &0.60$_{-0.08}^{+0.11}$ & 1.07$_{-0.15}^{+0.20}$ & 0.45$_{-0.06}^{+0.06}$ & 0.80$_{-0.11}^{+0.11}$ & 0.52$_{-0.05}^{+0.06}$ & 0.93$_{-0.10}^{+0.11}$ \\
       Ne & 1.42$_{-0.10}^{+0.10}$ & 1.73$_{-0.51}^{+0.61}$& 1.22$_{-0.37}^{+0.44}$ & 0.9$_{-0.34}^{+0.43}$ & 0.63$_{-0.24}^{+0.31}$ & 0.90$_{-0.41}^{+0.55}$ & 0.63$_{-0.29}^{+0.39}$ &
       0.55$_{-0.21}^{+0.23}$ & 0.38$_{-0.15}^{+0.16}$ \\
       Mg & 0.42$_{-0.05}^{+0.06}$ & 0.56$_{-0.22}^{+0.26}$& 1.33$_{-0.55}^{+0.65}$ & 1.0 & & 0.32$_{-0.16}^{+0.15}$ & 0.76$_{-0.39}^{+0.37}$ &
       0.23$_{-0.11}^{+0.12}$\ & 0.55$_{-0.27}^{+0.30}$ \\
       Si & 0.63$_{-0.06}^{+0.07}$ & 0.63$_{-0.16}^{+0.18}$& 1.0$_{-0.27}^{+0.31}$ &  0.24$_{-0.13}^{+0.14}$ & 0.39$_{-0.21}^{+0.23}$ & 0.42$_{-0.11}^{+0.11}$ & 0.67$_{-0.19}^{+0.19}$ &
       0.30$_{-0.08}^{+0.08}$ & 0.48$_{-0.13}^{+0.14}$ \\
       S & 0.41$_{-0.10}^{+0.10}$ & 0.65$_{-0.25}^{+0.25}$& &0.38$_{-0.29}^{+0.29}$ & & 0.38$_{-0.17}^{+0.17}$ & &
       0.50$_{-0.14}^{+0.14}$\\
       \hline
    &  & \multicolumn{2}{c}{Flare N8} & 
       \multicolumn{2}{c}{Flare N9} & \multicolumn{2}{c}{Flare N10} & \multicolumn{2}{c}{Flare N13} \\
       \hline
       & X/Fe & X/Fe & Flare/Quies. & X/Fe & Flare/Quies. & X/Fe & Flare/Quies. & X/Fe & Flare/Quies.  \\
       \hline
       O & 0.56$_{-0.02}^{+0.02}$ & 0.51$_{-0.07}^{+0.08}$ & 0.91$_{-0.13}^{+0.15}$ & 0.50$_{-0.06}^{+0.06}$ & 0.89$_{-0.11}^{+0.11}$& 0.59$_{-0.06}^{+0.06}$ &1.1$_{-0.11}^{+0.11}$ & 0.60$_{-0.11}^{+0.16}$ & 1.10$_{-0.20}^{+0.29}$ \\
       Ne & 1.42$_{-0.10}^{+0.10}$ & 1.49$_{-0.45}^{+0.64}$ & 1.1$_{-0.33}^{+0.46}$ & 1.39$_{-0.37}^{+0.30}$ & 0.98$_{-0.27}^{+0.22}$ & 0.64$_{-0.27}^{+0.29}$  & 0.45$_{-0.19}^{+0.21}$ & 2.1$_{-0.83}^{+0.68}$ &  1.5$_{-0.59}^{+0.49}$ \\
       Mg & 0.42$_{-0.05}^{+0.06}$ & 0.33$_{-0.17}^{+0.20}$ & 0.79$_{-0.42}^{+0.49}$ & 0.31$_{-0.18}^{+0.20}$& 0.74$_{-0.44}^{+0.49}$ & 0.34$_{-0.13}^{+0.14}$ & 0.81$_{-0.32}^{+0.35}$ &
       0.78$_{-0.54}^{+0.74}$ & 1.9$_{-1.3}^{+1.8}$ \\
       Si & 0.63$_{-0.06}^{+0.07}$ & 0.57$_{-0.16}^{+0.19}$ & 0.91$_{-0.27}^{+0.32}$ & 0.49$_{-0.17}^{+0.19}$ &0.78$_{-0.28}^{+0.31}$ & 0.50$_{-0.10}^{+0.11}$ & 0.79$_{-0.18}^{+0.20}$ & 
       0.71$_{-0.44}^{+0.54}$ & 1.12$_{-0.71}^{+0.86}$ \\
       S & 0.41$_{-0.10}^{+0.10}$ & 1.0 & & 1.0 & & 0.57$_{-0.16}^{+0.16}$ & &
       1.0 \\
       \hline
      \end{tabular}
\end{table*}
\label{table:abundances in NICER flares}
We list the abundances of five elements: Oxygen (O), Neon (Ne), Magnesium (Mg), Silicon (Si) and Sulfur (S) in Table \ref{table:abundances in NICER flares} obtained by fitting the spectrum (using \texttt{xspec}) of the quiescent level of the star, and those of the eight largest flares: N0, N2, N4, N5, N8, N9, N10 and N13. All the abundances are expressed relative to Iron (Fe). We were not able to fit the abundance of S in the quiescent level properly. So a default value of 1.0 is listed in the table. For each flare, we compare the abundances of four elements: O, Ne, Mg and Si with respect to the quiescent level. Such values are listed in the columns `Flare/Quies.' Using these ratios, we analyze the FIP/IFIP effect during the eight largest flares on EV Lac.  

In Figure \ref{fig:abundance ratio}, we plot the values of abundance ratios `Flare/Quies.' of four elements: O (black triangle), Ne (blue square), Mg (red circle) and Si (pink diamond) as a function of the FIP of elements. The values of FIP of O, Ne, Mg and Si are 13.62, 21.56, 7.65 and 8.15 eV respectively. So among these four elements, O and Ne are high-FIP elements and \textbf{the other two are} low-FIP elements. The dashed horizontal line in each subplot corresponds to the abundance ratio of the elements during quiescent state of the star. The flare label is mentioned inside each subplot. The plots suggest that low-FIP element Mg showed no significant change in abundance during seven flares except during flare N5. The large ratio for Mg during flare N2 is due to the fact that its real value could not be fitted and a default value equal to 1.0 was used during spectral fitting. Another low-FIP element Si also showed no significant change during four flares and is \textbf{found to be} under-abundant during the remaining four flares.  While O is found to be under-abundant \textbf{compared to} the quiescent value during one flare and over-abundant during one flare, Ne is found to be under-abundant \textbf{compared to} the quiescent value during three flares. In general, we cannot identify with confidence any overall patterns regarding the FIP effect in the EV Lac flares we have reported here. 
\begin{figure*} \label{fig:nicer lc}
   \centering
   \includegraphics[width=\textwidth,height=0.4\textwidth]{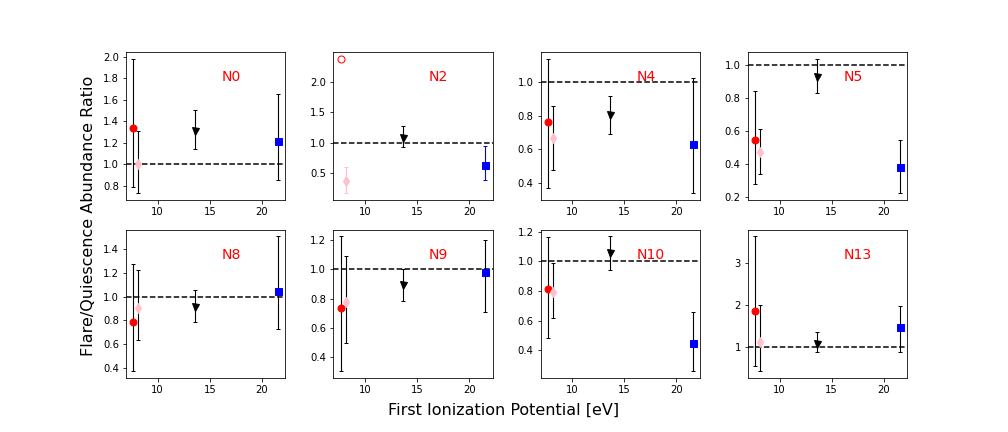}
    \caption{Flare abundance ratio for four elements: O, Ne, Mg and Si, as a function of FIP. \textbf{In each subplot, the red circle corresponds to Mg, the pink diamond corresponds to Si, the black triangle corresponds to O and the blue square corresponds to Ne.} The horizontal dashed line corresponds to the abundance ratio of the element during quiescent state of the star. \textbf{A default value of 1.0 is used to estimate the ratio for Mg in the case of flare N2. So a hollow circle is used in the corresponding plot.}}
\end{figure*}
\label{fig:abundance ratio}
\section{Summary and Discussion}
\label{sec:summary_discussion}
We acquired data of the nearby dM3.5e star EV Lac using 5 different observatories: NASA's TESS mission, NASA's Neil Gehrels Swift Observatory (\textit{Swift}), NASA's Neutron Interior Composition Explorer (NICER) and two ground based telescopes (University of Hawaii 2.2-m (UH88) and Las Cumbres Observatory Global Telescope (LCOGT) Network). Our goal was to carefully characterize an ensemble of flaring events observed simultaneously in different wavelengths to understand how flare energies and frequencies are related at different wavelengths. During the $\sim$24 days of continuous TESS Cycle 2 observations, we acquired 3 simultaneous 18 ks UV/X-ray observations using \textit{Swift}, 21 simultaneous 97.7 ks X-ray observations using NICER, $\sim$3.0 hrs of ground-based observation with UH88, and $\sim$3.0 hrs of ground-based observation with LCOGT. 

We identified 56 white light flares in the TESS light curve, with estimated energies in the range log $E_{\rm T}$ (erg) = (30.5 - 33.2). 9 UV flares were identified in the \textit{Swift}/UVOT light curve, with estimated energies in the range log $E_{UV}$ (erg) = (29.3 - 31.1), but 3 were not observed throughout their full duration. Likewise, we identified  14 X-ray flares in the NICER light curve, with estimated minimum energies in the range log $E_{N}$ (erg) = (30.5 - 32.3). One flare with an estimated energy log $E_{L}$ (erg) = 31.6 was identified in the LCOGT light curve. One flare was observed simultaneously by TESS, UVOT, and NICER during various phases. While UVOT observed the rise and initial decay phase, NICER observed the later part of the decay phase. NICER and UVOT observed different phases of another flare. Likewise, TESS and NICER observed nine flares simultaneously. We did not identify any flares in the \textit{Swift}/XRT or UH88 light curves. 

In general, we might expect to observe flares in X-rays and UV simultaneously. But this was not the case during the observation by \textit{Swift}. However, it is also possible to see UV flares without X-ray flares since the two flare signatures are formed in different parts of the stellar atmosphere: the lack of correspondence implies that energy release is happening so low in the atmosphere that the chromospheric response (detected by \textit{Swift}/UVOT) is dominant. The lack of an X-ray flare implies that the corona is not involved in the event, perhaps because of a deficiency in the amount of evaporated chromospheric material. In contrast, in the standard scenario of a flare, the energy input from magnetic reconnection and accelerated particles heats chromospheric material on a timescale that is short compared with the hydrodynamic expansion time, causing the ablation of chromospheric material (now heated to coronal temperatures) up the loop legs which produces X-ray radiation. 

In order to ascertain the physical properties of the X-ray flares, we fitted the NICER flare spectra using a three-discrete temperature (3$T$) plasma model in Section \ref{subsection:NICER analysis}. The temperatures ($T_{1}$) of the coolest component ranges from 2.8 to 5.7 MK. We interpret the quiescent (i.e., non-flaring) corona as the origin of the $T_{1}$ component. Flares are expected to be confined to a relatively small area on the stellar surface, of order 1\% for white light flares and perhaps as large as 10\% for the Balmer line-emitting region (see \citealt{1982ApJ...257..269C}). EV Lac's quiet corona likely has temperatures of a few million degrees and might occupy 90\% or more of the surface area during the time of the flare. In partial support of this claim, we note that \cite{2006ApJ...647.1349O} estimated the differential emission measure (DEM) for the quiescent atmosphere of EV Lac and they found that it peaks at $T$ = 2.5 MK which overlaps within the error bars of the values of $T_{1}$ listed in Table \ref{table:NICER XRT} above. The temperatures ($T_{2}$) of the second component have values ranging from 7.5 to 11.5 MK. We interpret these temperatures as being due to a hotter component of the quiescent corona in EV Lac. \cite{2006ApJ...647.1349O} also found a secondary peak in the DEM for a quiescent EV Lac at $T$ = 7.9 MK, which overlaps with the range of values we have obtained from NICER data for the quantity $T_{2}$. We find the temperatures $T_{3}$ of the third component have values from 22.0-29.0 MK and correspond to the flaring corona. \cite{2005ApJ...621..398O} analyzed nine X-ray flares observed by $Chandra$ with a 2$T$ plasma model and found that for all but two flares, the hotter (i.e. flaring) component had values in a range similar to our $T_3$ components, lending confidence to our interpretation. 

Taking into account the minimum X-ray flare energies observed by NICER, we estimated the maximum energy ratio of optical and X-ray flares observed by TESS and NICER. We find that the ratio exceeds unity for almost all flares, and by an order of magnitude for the larger flares. These results are to some extent consistent with the results of \cite{2019A&A...628A..79S} who analyzed a sample of 8 superflares (with energies $\geq$10$^{34}$ ergs) on another flare star (AB Doradus), which were detected by TESS in the course of about two months of observations. The TESS energies ranged from 1-50 $\times$ 10$^{34}$ ergs. They also compared the energies of those flares to those of 34 X-ray flares (with energies in the range log $E$ = 30.03 to 33.83) observed on the same star by XMM-Newton in a period of 11 years \citep{2016IAUS..320..155L} and to that of the largest solar flare seen in solar irradiance measurements. They found, on average, the total X-ray energy of flares on AB Dor to be less than the energy in optical (super)flares. In an earlier paper, \cite{1976ApJ...207..289M} used the values of time scales of radiative energy loss and conductive energy loss to estimate the ratio of luminosities in X-rays and in optical photons: $L_{\rm X}$/$L_{\rm opt}$. He found that $L_{\rm X}$/$L_{\rm opt}$ would be no larger than 0.1 (sometimes considerably less than 0.1). The conclusions of \cite{2019A&A...628A..79S} and \cite{1976ApJ...207..289M} suggest that the ratios of flare energies we have estimated in Table \ref{table:tess_nicer energy comparison} may not be significantly different from the real values. 

We searched for the FIP/IFIP effect using the abundances of four elements (O, Ne, Mg and Si) during the eight largest NICER flares. We find that two elements Ne (high FIP element) and Si (low FIP element) both are under-abundant relative to the quiescent state during three flares, and Si is under-abundant in one more flare. In an exceptional flare, Ne was found to be under-abundant relative to the quiescent state. \textbf{The next highest FIP element studied, O,} was found to be under-abundant in one of the NICER flares and over-abundant in another flare. The low-FIP element Mg was neither over-abundant nor under-abundant relative to the quiescent state in all flares except one. Thus, we cannot draw any definite conclusions regarding a pattern of either FIP or IFIP during the EV Lac NICER flares.

Oscillatory and pulsating signatures known as quasi-periodic pulsations (QPPs) are a common feature observed in the light curves of both solar and stellar flares (e.g. \cite{Vida2019} (TESS), \cite{Pugh2016} (\emph{Kepler}),  \cite{Jackman2019} (NGTS), \cite{Broomhall2019} (XMM-Newton), \cite{Inglis2016} (in solar flares)). These oscillations can provide constraints on the mechanisms of flare production and the properties of the stellar atmosphere. This includes: thermal free-free microwave emission of chromospheric plasma heated during a flare and filling in the flaring loop \citep{Kupriyanova2014}, modulation of the non-thermal electron dynamics by MHD oscillations \citep{Zaitsev1982} and the triggering of energy released by external MHD waves or oscillations  \citep{Nakariakov2016,Nakariakov2006,Chen2006}. QPP periods can range from a fraction of a second to several minutes \citep{Nakariakov2016,VanDoorsselaere2016}. The short lived nature and short periods of QPPs makes the $Swift$/UVOT and NICER data presented here, with cadences of $\sim$10 seconds, ideal for a follow-up study to search for and analyze QPPs in stellar flare morphology. We will present our results regarding QPPs in EV Lac flares in Monsue et al. 2021 (in prep.).

\subsection{Comparison of flare frequency distributions}
For each of the three spectral bands in which we have observed multiple flares (TESS, $Swift$/UVOT and NICER), we have used least-squares fitting to obtain a flare frequency distribution described by power law index $\beta$ (Table \ref{table:power law fit}). We used Poisson uncertainties to weight the fit which helps to minimize the bias due to potential over-weighting of the largest flare energies.  The values of minimum and maximum energies, and the number of flares used for fitting are also listed. The errors are obtained by dividing the values of $\beta$ by $\sqrt{N}$, where $N$ is the number of flares used for fitting. 

\begin{figure}
   \centering
   \includegraphics[width=0.5\textwidth,height=0.35\textwidth]{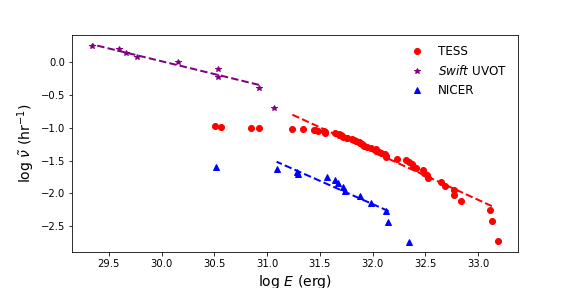}
    \caption{Comparison of FFDs of flares observed by TESS (red), NICER (blue), and \textit{Swift}/UVOT (purple). The dashed lines is power-law fit to the distribution of flares in a given band.}
    \label{fig:ffd_comparison}
\end{figure}

In Figure \ref{fig:ffd_comparison}, we compare the FFDs of flares observed by TESS, NICER and \textit{Swift}/UVOT. The X-ray FFD in this figure is estimated by using the same observation time (i.e., 23.2 d) as TESS.  \footnote{Since NICER observations were performed throughout a single TESS sector ($\sim$25) days, using the total exposure time ($\sim$98 ks) of NICER to estimate the FFD would result in flare rates higher than the real values.} We remind the readers that the FFD of NICER data is estimated by using the lower limit of all flare energies as none of the flares were observed for the full duration. It is interesting to note that the slopes of the FFDs corresponding to TESS and NICER data are comparable. This might be due to the fact that the flares observed on EV Lac by TESS and NICER were emitted as the ultimate result of a common physical phenomenon (e.g. non-thermal electrons). Previously, \cite{2015ApJ...809...79O} also found that correcting for the fraction of the bolometric flare energy released in the different wavebands (coronal, optical), the optical FFD and coronal FFDs were consistent with each other, indicating a continuation of a common trend over a wide range of flare energies. 

Since the $Swift$/UVOT flares U2, U7 and U8 were not observed for their full duration, the values of energies of those flares do not represent the total energies of corresponding flares. The largest energy (i.e., log E (erg) $>$ \textbf{33.3}, in Table \ref{table:properties of UVOT flares}) is not included in the FFD fit to reduce bias. The reported energy of flare U2, the next largest flare, represents almost all the released energy since only a short part of U2's decay phase was not observed. So we claim that the FFD of UVOT flares estimated in this paper is a good representation of flare rates even though the number of flares is small. Though we have a small number of UV flares, the shallower FFD slope matches the results of \cite{2018ApJ...854...14M} who found that for a given star, the FFD turns over to shallower slopes for low energy flares in the log $\tilde{\nu}$ versus log $E$ diagram. For a slightly different flare energy range than that reported in this paper, \cite{1976ApJS...30...85L} reported a value of $\beta$ = -0.69$\pm$0.11 for EV Lac for U-band energies in the range log $E_{U}$ $\sim$(30.5 - 32.5). However, they also included the energies of flares observed in other filters (B and V), which were converted to U-band energies. The value of $\beta$ also depends on the range of flare energies used for fitting. 

\cite{2000ApJ...541..396A} used $EUVE$ data to study the distribution of coronal (EUV and X-ray) flare rates of EV Lac. They estimated $\beta$ = 0.76$\pm$0.33 which matches well with our power-law fit to NICER X-ray flare energy distribution within 1-$\sigma$. Likewise, \cite{1988A&A...205..197C} estimated $\beta$ = 0.52 for soft X-ray flares on M dwarfs by using $EXOSAT$ data. Their results also agree with ours within 1-$\sigma$. 

%
\begin{table}
 	\caption{Power law fit to FFDs}
 	\label{table:power law fit}
     \centering
     \begin{tabular}{ccccc}
     \hline
       Band & $\beta$ & log $E_{\rm min}$ & log $E_{\rm max}$ & \# of flares \\
       \hline
        & & erg & erg \\
       \hline
       TESS & -0.67$\pm$0.09 & 31.2 & 33.1 & 51 \\
       \textit{Swift}/UVOT & -0.38$\pm$0.13 & 28.7 & 30.2 & 8 \\
       NICER & -0.65$\pm$0.19 & 31.1 & 32.2 & 12 \\
     \hline
      \end{tabular}
\end{table}
\section{conclusions} In this work we present simultaneous multiwavelength flare observations of the nearby, active M dwarf EV Lac. We obtained time series observations in the optical, UV, and X-ray and analyze the star's flares in each band and use these results to compare the multiwavelength flare properties. Our major findings are:
\begin{itemize}
\setlength\itemsep{-0.1em}
    \item TESS data reveal EV Lac has WLF rate of $\sim$2 per day and a rotation period of 4.3592 days. 
    \item {\it Swift} UVOT observations indicate that EV Lac has a higher flare rate in the UV than in white light or X-ray, although this may be because the flares detected with UVOT are all of significantly smaller equivalent duration and energy than the TESS and NICER flares. The low energy {\it Swift} UV flares also have larger amplitudes compared to the white light flares, likely because the flare emission peaks in the UV.
    
    \item Our results (in Table \ref{table:tess_nicer energy comparison}) suggest that the flare luminosity in the optical is comparable to (or somewhat larger than) the flare luminosity in X-rays. This feature of flare radiation could be consistent with thermal processes of energy distribution (conduction, radiation) in flare plasma. This feature could also be consistent with bremsstrahlung emission from hot flare plasma \citep{1965IrAJ....7...20A, 1976ApJ...210..702M}: see especially \cite{1977A&A....61..625K} for flares in EV Lac in particular. Bremsstrahlung emission extends at essentially constant flux at all frequencies which are less than h$\nu_{\rm max}$ $\approx$ kT. With $T$ = 3-30 MK in flare plasma (see Table \ref{table:NICER XRT}), such a spectrum could account for comparable energies radiated in $\sim$1 keV X-rays and in optical light. However, it is not clear that a single mechanism can explain the various emissions which have been identified in stellar flares \citep{2013ApJS..207...15K}: these include a black-body continuum with a temperature of order 10$^{4}$ K, emission in the Balmer continuum and in high-level Balmer lines, and a mysterious “conundrum” which appears at wavelengths redward of 6000 \AA  (possibly due to H-emission). Until such times as the contributions from these various components can be simultaneously quantified, it will be difficult to state definitively how flare energy is partitioned across the spectrum.
    \item \textbf{Given the uncertainties of the fitted abundances of four elements (O, Ne, Mg and Si), the current study could not conclusively find evidence of neither the FIP nor IFIP effect during the flares observed on EV Lac.} \cite{2015LRSP...12....2L} has proposed a model for generating the FIP and the IFIP in partially ionized plasma, such as occur in the chromosphere of all cool stars. MHD waves propagating through the chromosphere exert a ponderomotive force on the ions but not (directly) on the neutrals, leading to ion-neutral fractionation. The fractionation has maximum amplitude at a certain altitude $H$ ($\approx$2150 km in the Sun). Laming shows that \textit{upward} propagating Alfven waves favor the creation of the FIP effect, whereas downward propagating fast-mode MHD waves favor the IFIP effect. Although the Sun in general displays the FIP effect in active regions, transient detection of IFIP has been reported occasionally in localized regions during flares \citep{2019ApJ...875...35B}. The geometry of the field (closed or open field lines?) also contributes to the FIP/IFIP effect, as does the altitude $H$ of predominant ion-neutral fractionation. The lack of an observed FIP or IFIP in EV Lac could be due to one or more of the following:   (i) the presence of complicated field topology; (ii) a mixture of upward and downward wave fluxes; (iii) an unfavorable location of the \textbf{altitude} $H$.
\end{itemize}

In conclusion, our multiwavelength study of flares in one particular flare star (EV Lac) has helped to confirm certain aspects as to how the radiant energy of flares is distributed across various regions of the electromagnetic spectrum. However, if it turns out that non-thermal electrons contain much of the flare energy (up to 50\% in a sample of “small” solar flares: e.g. \citealt{1971SoPh...17..412L}), then it could be beneficial to include, in any future multiwavelength study of flares, observations of non-thermal X-rays. A flare-associated population of non-thermal electrons might in principle also be tracked by means of radio observations, and EV Lac is already known to emit circularly polarized flaring radiation at centimeter wavelengths \citep{2005ApJ...621..398O}. Unfortunately, Osten et al reported that there seems to be no obvious relationship \textit{in the timing} between the flares which they detected in centimeter radio, X-ray, or optical. In a joint optical-radio study of flaring stars \textbf{\citet{1976ApJ...203..497S}} reported a certain correlation in $timing$ between the light curves in radio and optical: the radio peak at 318 MHz was found on average to be $delayed$ by 0-5 minutes relative to the optical peak. If this time-delay in the light curves is related to physical processes in the flare and/or in the corona, then future multi-wavelength campaigns could benefit from the use of meter-wave radio data.  

\acknowledgements
The material in this paper is based upon work supported by NASA under award No. 80NSSC19K0104, 80NSSC19K0315 and 80GSFC21M0002. D.H. acknowledges support from the Alfred P. Sloan Foundation, the National Aeronautics and Space Administration (80NSSC19K0379), and the National Science Foundation (AST-1717000). M.A.T. acknowledges support from the DOE CSGF program through grant DE-SC0019323. This paper includes data collected by the TESS mission. Funding for the TESS mission is provided by the NASA Explorer Program. This work made use of data supplied by the UK Swift Science Data Centre at the University of Leicester. This research has made use of the XRT Data Analysis Software (XRTDAS) developed under the responsibility of the ASI Science Data Center (ASDC), Italy. This work makes use of observations from the LCOGT network. This research has made use of data obtained through the High Energy Astrophysics Science Archive Research Center Online Service, provided by the NASA/Goddard Space Flight Center. This publication makes use of data products from the Wide-field Infrared Survey Explorer, which is a joint project of the University of California, Los Angeles, and the Jet Propulsion Laboratory/California Institute of Technology, funded by the National Aeronautics and Space Administration. This publication makes use of data products from the Two Micron All Sky Survey, which is a joint project of the University of Massachusetts and the Infrared Processing and Analysis Center/California Institute of Technology, funded by the National Aeronautics and Space Administration and the National Science Foundation. This research made use of Photutils, an Astropy package for
detection and photometry of astronomical sources.

The authors wish to recognize and acknowledge the very significant cultural role and reverence that the summit of Maunakea has always had within the indigenous Hawaiian community.  We are most fortunate to have the opportunity to conduct observations from this mountain.

\software: Astropy \citep{astropy:2013,astropy:2018}, Matplotlib \citep{Hunter:2007}, Numpy \citep{2020NumPy-Array}, Lightkurve \citep*{lightkurve}, Python, IPython \cite{2007CSE.....9c..21P}, Jupyter \citep{Kluyver:2016aa}
\facility: 
TESS, \textit{Swift} XRT/UVOT, NICER \\
University of Hawaii 2.2 m Telescope (UH88), Las Cumbres Observatory Global Telescope (LCOGT)
\bibliographystyle{aasjournal}
\bibliography{astrobib}

 \end{document}